\documentclass[12pt]{iopart}
\usepackage{graphicx,epstopdf}
\usepackage{caption}
\usepackage{subcaption}
\usepackage{lscape}
\usepackage{hyperref}
\usepackage{microtype}
\graphicspath{ {./figs/} }
\begin{document}

\title[Neural Co-Processors]
{Neural Co-Processors for Restoring Brain Function:\\Results from a Cortical Model of Grasping}

\author{Matthew J Bryan$^{1}$, Linxing Preston Jiang$^{1,2,3}$, Rajesh P N Rao$^{1,2,3}$}

\address{$^{1}$ Neural Systems Laboratory, Paul G. Allen School of Computer
Science \& Engineering, University of Washington, Seattle, WA, USA}
\address{$^{2}$ Center for Neurotechnology, University of Washington, Seattle, WA, USA}
\address{$^{3}$ Computational Neuroscience Center, University of Washington, Seattle, WA, USA}

\ead{\{mmattb,prestonj,rao\}@cs.washington.edu}
\vspace{10pt}
\begin{indented}
\item[]March 2023
\end{indented}

\begin{abstract}
\textit{Objective} A major challenge in designing closed-loop brain-computer interfaces (BCIs)
is finding optimal stimulation patterns as a function of ongoing neural activity for
different subjects and different objectives. Traditional approaches, such as
those currently used for deep brain stimulation (DBS), have largely followed a manual
trial-and-error strategy to search for effective open-loop stimulation parameters,
a strategy that is inefficient and does not generalize to closed-loop
activity-dependent stimulation.
\textit{Approach} To achieve goal-directed closed-loop neurostimulation, we propose
the use of brain co-processors, devices which exploit artificial intelligence (AI)
to shape neural activity and bridge injured neural circuits for targeted repair and
restoration of function.  Here we investigate a specific type of co-processor called a ``neural
co-processor'' which uses artificial neural networks (ANNs) and deep learning to learn optimal
closed-loop stimulation policies. The co-processor adapts the stimulation policy as
the biological circuit itself adapts to the stimulation, achieving a form of brain-device
co-adaptation. Here we use simulations to lay the groundwork for future \textit{in vivo}
tests of neural co-processors. We leverage a previously published cortical model of grasping, to
which we applied various forms of simulated lesions. We used our simulations to develop
the critical learning algorithms and study adaptations to non-stationarity in preparation for
future \textit{in vivo} tests.
\textit{Main results} Our simulations show the ability of a neural co-processor to learn a
stimulation policy using a supervised learning approach, and to adapt that policy as
the underlying brain and sensors change. Our co-processor successfully co-adapted with
the simulated brain to accomplish the reach-and-grasp task after a variety of lesions
were applied, achieving recovery towards healthy function in the range 75-90\%.
\textit{Significance} Our results provide the first proof-of-concept demonstration,
using computer simulations, of a neural co-processor for adaptive activity-dependent closed-loop
neurostimulation for optimizing a rehabilitation goal after injury. While a significant
gap remains between simulations and \textit{in vivo} applications, our results provide
insights on how such co-processors may eventually be developed for learning complex
adaptive stimulation policies for a variety of neural rehabilitation and neuroprosthetic
applications. 
\end{abstract}

\vspace{2pc}
\noindent{\it Keywords}: brain-computer interface, brain-machine interface, neurostimulation,
neuromodulation, neural co-processor, AI, machine learning, deep learning, neural networks,
computational models
%
%
\maketitle
%
%

\section{Introduction}
Brain-computer interfaces (BCIs) have made significant advances over the last several decades, leading
to the control of a wide variety of virtual and physical prostheses through neural signal decoding
\cite{rao.bcibook, wolpaw.bcibook, moritz.neuro, lebedev.bmi}. Separately, advances in stimulation
techniques and modeling have allowed us to probe neural circuit dynamics (e.g. \cite{walker.inception})
and learn to better drive neural circuits towards desired target dynamics by encoding and delivering
information through stimulation \cite{niparko.cochlear, weiland.retinal, tomlinson.propr, tabot.tact,
tyler.tact, dadarlat.tact, sharlene.tact, cronin.tact}. Bi-directional BCIs (BBCIs) allow stimulation
to be conditioned on decoded brain activity and encoded sensor data for applications such as real-time,
fine-grained control of neural circuits and prosthetic devices (e.g., \cite{nicolelis.bmbi}).

Motivated by these advances, we investigate here a flexible framework for combining encoding
and decoding using ``neural co-processors'' \cite{rao.coproc}, a type of brain co-processor \cite{rao.braincoproc}.
Neural co-processors leverage artificial neural networks (ANNs) and deep
learning to compute optimal closed-loop stimulation patterns. The approach can be used to not only drive neural activity
toward desired activity regimes, but also to achieve task goals external to the subject, such as finding closed-loop
stimulation patterns for motor cortical neurons for restoring the ability to reach and grasp an object. Likewise, the
framework generalizes to stimulation based on both brain activity and external sensor measurements, e.g., from cameras or
light detection and ranging (LIDAR) sensors, in order to restore perception (e.g., cortical visual prosthesis) or
incorporate feedback for real-time prosthetic control (see \cite{rao.braincoproc} for details).

The co-processor framework also allows co-adaptation with biological circuits in the brain by updating its 
stimulation policy, while the brain updates its own response to the stimulation via adaptation and neural
plasticity, or modifies its response due to other reasons. The co-processor could potentially optimize its outputs 
for a desired optimization function continually in the presence of significant non-stationarities in the brain.

Here, to lay the groundwork for future \textit{in vivo} tests of the co-processor framework, we 
use computer simulations to explore how co-processors can be trained to restore lost function and how they can adapt to 
 non-stationarities. We demonstrate a neural co-processor that restores movement in a
computational model of cortical networks involved in controlling a limb, after a simulated stroke affects the
ability to use that limb. Our demonstration combines both components of a neural co-processor \cite{rao.coproc}:
\begin{itemize}
	\item An emulation model based on ANNs, which learns a mapping from  
	      stimulation and current neural activity to output variables such as task performance (or future neural activity).
	\item An artificial intelligence (AI) ``agent'' based on ANNs which learns the best  closed-loop
          activity-dependent stimulation to apply in real time to optimize a given task.
\end{itemize}

\section{Background}
\label{sec:background}

Significant advances have been made in understanding and modeling the effects of electrical and other forms of
neural stimulation on the brain. Researchers have explored how information can be biomimetically or
artificially encoded and delivered via stimulation to neuronal networks in the brain and
other regions of the nervous system for auditory \cite{niparko.cochlear}, visual \cite{weiland.retinal},
proprioceptive \cite{tomlinson.propr}, and tactile
\cite{tabot.tact, tyler.tact, dadarlat.tact, sharlene.tact, cronin.tact} perception.
Advances have also been made in modeling the effects of stimulation over large scale, multi-region
networks, and across time \cite{shanechi.stimmodel}. Some models can additionally adapt to ongoing
changes in the brain, including changes due to the stimulation itself
\cite{tafazoli.acls}. For our simulations described below, we use a stimulation
model, not unlike those cited above, which seeks to account for both network dynamics
and non-stationarity. In addition to training the model to have a strong ability to predict
the effect of stimulation, we additionally adapt it to be useful for learning an
optimal stimulation policy, a property distinct from predictive
power alone.

Researchers have also explored both open- and closed-loop stimulation protocols for
treating a variety of disorders. Open loop stimulation has been effective in
treating Parkinson's Disease (PD) \cite{benabid.parkinsons}, as well as various
psychiatric disorders \cite{holtzheimer.psy, kisely.psy, fraint.psy}.
In research more directly related to our work, Khanna et al. \cite{khanna.openloop}
investigated the use of open loop stimulation in restoring dexterity after a lesion
in nonhuman primate's (NHP) motor cortex. The authors demonstrate that the use
of low-frequency alternating current, applied epidurally, can improve grasp performance.

While open loop stimulation techniques have yielded clinically useful results, results
in many domains have been mixed, such as in visual prostheses \cite{bosking.visual},
and in invoking somatosensory feedback \cite{cronin.tact}. We believe this is due to the
stimulation not being conditioned on the ongoing dynamics of the neural circuit being
stimulated. From moment to moment and throughout the day, a neuronal circuit in the brain
can be expected to respond differently even when the same stimulation parameters are used,
due to the multitude of different external and internal inputs influencing the circuit's
ongoing activity. Stimulation therefore needs to be closed-loop, i.e. proactively adapted
in response. This need is even greater over longer time scales as the effects of plasticity,
changes in clinical conditions, and ageing change the dynamics and connectivity of the brain.
Closed-loop stimulation may also provide means to better regulate the energy use of an
implanted stimulator, allowing it to intelligently regulate when to apply stimulation,
in order to preserve implant battery life \cite{castano.pd}. Another benefit is that closed-loop
stimulation offers an opportunity to minimize the side-effects of stimulation, through real
time regulation of the stimulation parameters, such as in the use of deep brain stimulation
(DBS) in PD patients \cite{little.park}. In recent years, closed-loop stimulation has been
used to aid in learning new memories after some impairment \cite{berger.closedloop,
kahana.biomarker}, to replay visually-invoked activations \cite{tafazoli.acls}, and for
optogenetic control of a thalamocortical circuit \cite{bolus.opto}, among others.

A major open question is: how does one leverage closed-loop stimulation for real-time
co-adaptation with the brain to accomplish an external task such as restoration of a lost function?
``Co-adaptation'' here refers to the ability of a BCI to adapt its stimulation regime to ongoing
changes in neural circuits in the brain, and to adapt with the brain to accomplish
the external task (e.g., grasping). The neural co-processor we present here provides one
potential approach to accomplishing this goal. Through the use of deep learning, a neural 
co-processor co-adapts its AI, which controls stimulation, in synch with the biological circuits in the brain.

For a neurologically complex task such as grasping, it is unlikely that there exists
a fixed real-time controller which can be identified \textit{a priori} for stimulating the (potentially
impaired) neural circuits involved in the task. This is due in large part to the variability
in the placement and performance of sensors and stimulators in different brains, as well as variability
in brain structure and function between subjects. The most plausible path to implementing a real-time
controller is therefore to allow the device to adapt to the subject, and to the long-term changes in
their brain activities, and variability in the sensors, stimulators and hardware. Our proposed neural
co-processors seek to accomplish such adaptation through ANNs and deep learning, and a particular
training paradigm described below.

\subsection{Simulation as a Way to Gain Insights Prior to {\em in vivo} Experiments}
To gain insights into neural co-processors before testing them \textit{in vivo}, we
investigated a number of crucial design elements through the use of a previously published model by
Michaels et al. \cite{michaels.mrnn} of the cortical areas involved in grasping; the model, based on
multiple recurrent neural networks, is inspired by cortical anatomy and was fit to data from nonhuman
primates performing grasping tasks. Using this cortical model as a ``simulated brain'' allowed us to
rapidly iterate through different design and training methods to demonstrate key properties of the
neural co-processor framework. These insights will help guide the co-processor training methodologies
and experimental design for future \textit{in vivo} experiments. Additionally, a
commonly-accepted maxim of animal experimentation is that animal use should be narrowly tailored
to answering questions which cannot be answered in other ways. Consider, for example, ``the 3Rs
alternatives'' approach to the use of animal experiments \cite{nrc.care}. In our case, we leverage
computer simulations for initial investigations into co-processor design, gathering evidence in
preparation for future \textit{in vivo} experiments. In the Discussion section, we explore
potential paths for translation of this work to \textit{in vivo} experiments.

A previous example of such a simulation approach is the work of Dura-Bernal et al.
\cite{bernal.sim}. In this work, the authors used a simulated spiking neural network
to train a stimulation agent. Their stimulation agent sought to restore the network's
control of a simulated arm to reach a target, after a simulated lesion was
applied. Similar to our approach, the authors simulated lesions by effectively removing parts
of their simulated network, or by removing connections between parts of the network.
As the authors point out, there exists only limited ability to probe a neural circuit
\textit{in vivo} in order to perform learning. As a result, we first need to design
our approach through the use of an admissible simulation. The key question then is:
how can we design an admissible simulation, i.e. one which provides a test bed on
which we can demonstrate key properties of our co-processor designs? We answer this question in our
work by adopting a a previously published model of cortical grasping \cite{michaels.mrnn} that has
been shown to replicate properties of the cortical circuit in the biological brain based on fits to
nonhuman primate data.

Through our simulation, we explored what properties of the co-processor allow successful
adaptation to the short-term dynamics of the cortical model as it is being stimulated,
as well as adaptation to longer-term connectivity changes in the cortical model.
We present a training method for neural co-processors for learning optimal stimulation
patterns that drive improvements in external task performance, while also adapting to the
non-stationarity of the stimulated neural circuits.

\section{Methods}

\begin{figure}
    \centering
    \begin{subfigure}[c]{0.88\textwidth}
		\centering
		\includegraphics[width=\textwidth]{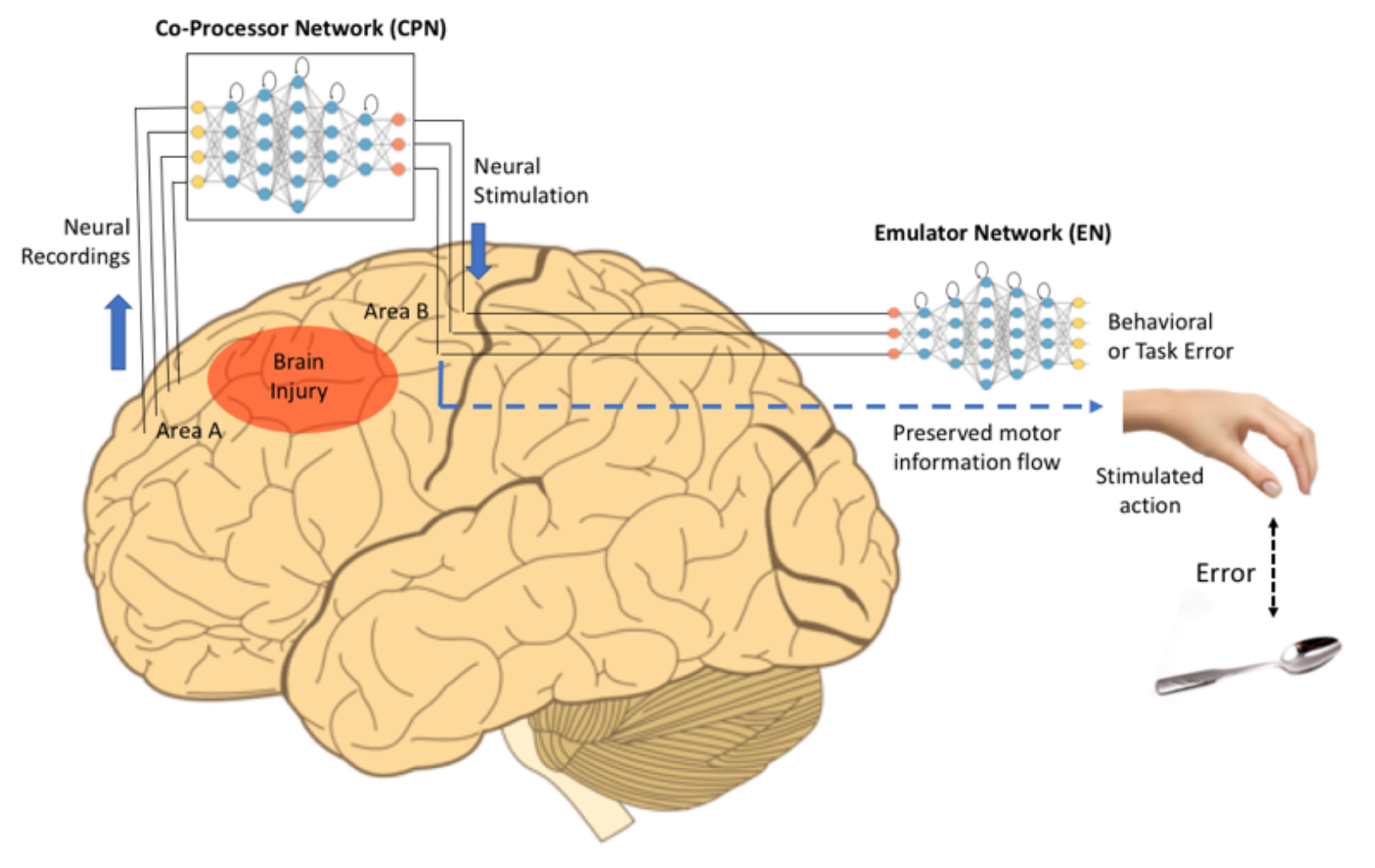}
		\caption{}
	\end{subfigure}
	\hfill
    \begin{subfigure}[c]{0.88\textwidth}
		\centering
		\includegraphics[width=\textwidth]{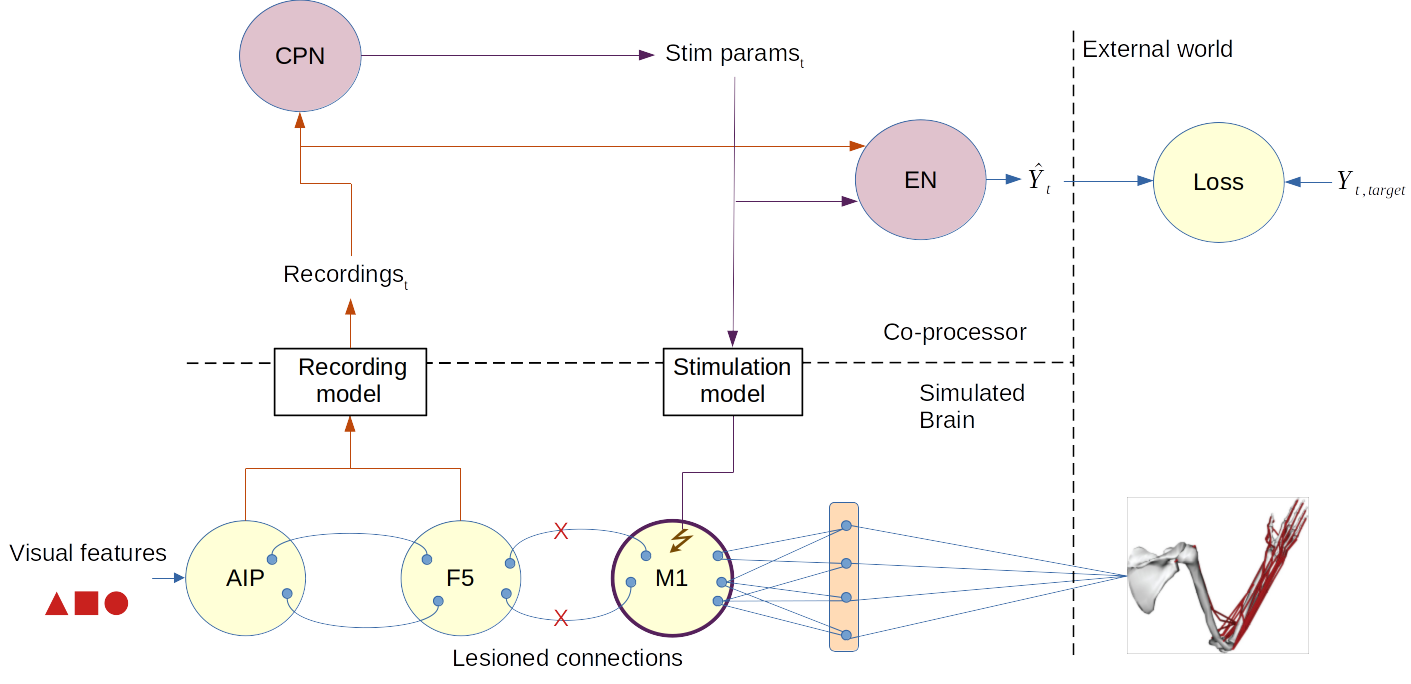}
		\caption{}
	\end{subfigure}
	\hfill
    \caption{{\bf Neural co-processor for restoring function after a brain injury}.
    An artificial neural network called the “Co-Processor Network” (CPN) is used to
    map input neural activity patterns in an area A to output stimulation patterns
    in same or other areas B in order to achieve a neural or behavioral goal using
    another ANN, an “Emulator Network” (EN) - see text for details. (a) The example
    here shows the CPN creating a new information processing pathway between prefrontal
    cortex and motor cortex, bypassing an intermediate area affected by brain injury
    (e.g., stroke). Adapted from \cite{rao.coproc}, (b) Our current study involves a
    simulated cortical grasping circuit. The CPN and EN both receive simulated brain
    recordings, which are created according to a recording model (see Methods).
    The CPN outputs stimulation parameters, which are applied to the simulated
    grasping circuit, according to a stimulation model. The EN models the relationship between
    the     stimulation, brain recordings, and external task.}
    \label{fig:arch}
\end{figure}

\subsection{Architecture Overview}
First, we present the architecture of our neural co-processor design. This design aims to solve two
fundamental challenges in using neural stimulation to improve external task performance.
First, to restore function for complex tasks (such as grasping), it is difficult to determine
the mapping between observed neural activity and the stimulation patterns to be applied to a downstream
motor area. As a result, the co-processor must learn what stimulation pattern is appropriate for achieving the
external task given the current neural activity. Unfortunately, in most cases, we do not know what the correct
stimulation pattern is for any given input activity pattern, a precondition for using supervised deep learning
methods for training the co-processor. Stimulation shapes the nonlinear dynamics of circuits in the brain
in complex ways, leading to complex effects in external behavior. It is therefore not obvious what
stimulation patterns will produce a desired behavior. 

A neural co-processor attempts to solve these problems with a pair of artificial neural networks 
(Fig.~\ref{fig:arch}):
\begin{itemize}
    \item a ``Co-Processor Network'' (CPN), which is a recurrent neural network (RNN) which maps neural activity, and possibly
	      data from external sensors, to appropriate stimulation parameters.
    \item an ``Emulator Network'' (EN), also an RNN which models the effect of stimulation on neural dynamics and
          behavior for the external task.
\end{itemize}

The CPN can be trained using the backpropagation algorithm, the workhorse of deep learning for training ANNs.
However, backpropagation requires the error between the output of the CPN and a desired output, and as
discussed above, we do not have the desired output stimulation pattern. We do however know what the desired
output behavior in a task should be, e.g., a particular kind of grasp for a particular object or a particular
type of neural activity in a brain area corresponding to healthy activity. We can therefore compute the error
between this desired behavior and the actual behavior caused by the CPN due to stimulation. How do we
backpropagate this external task error to update the parameters of the CPN so that it learns to produce
stimulation patterns that are optimal for the task?

We use a trained EN to backpropagate the external task error to the CPN, which then uses this backpropagated
error to update its parameters. We train the EN to predict task-relevant parameters - a prediction of muscle
velocities in our case - given randomly sampled stimulation patterns or stimulation
patterns output by the CPN, and any measured neural activity. If the EN is trained to a sufficiently
high level of precision, it can be used as a function approximator for the subject's true ``stimulation function''.
When training the CPN, we treat the EN's output as the actual task output (e.g., actual grasp
behavior or muscle velocities). We backpropagate through the EN (without changing its parameters) and then the
CPN (changing its parameters) the error between the EN's output and the desired task output (see Fig. \ref{fig:arch}). 

In our experiments, the EN was a single-layer fully-connected long short-term memory (LSTM) recurrent
neural network, with hyperbolic tangent ($tanh$) activations, and a linear readout. It had 87 LSTM neurons,
the number being chosen somewhat arbitrarily as a function of the input and output vector sizes. We found that
varying this neuron count did not drastically change results. Although other architectures could also be used,
we found that this LSTM architecture allows the EN to continuously adapt to long-running dependencies in the
simulated neural dynamics, far better than a vanilla RNN. The CPN had an almost identical architecture, but with
61 LSTM neurons. As with the CPN, this neuron count was chosen as a function of the input and output vector sizes,
and increasing it had little effect on results. There is no requirement for the EN and CPN to have similar network
architectures, but we found that these choices worked well in our experiments.

Note that the EN is more general than traditional models of neurostimulation which attempt to predict the effects of
stimulation on { \em neural activity}. The EN in a neural co-processor predicts the effects of stimulation (taking into
account ongoing neural dynamics) on {\em task performance}. This provides the key functionality needed to train the
CPN. In the special case where the task involves driving neural circuits in the brain to desired neural activities,
the EN reduces to more traditional models of stimulation.

For comparison, consider an EN architecture based on a traditional RNN with a nonlinearity. This is a nonlinear version
of the common linear time-invariant state space stimulation model studied in previous research (see, e.g., \cite{shanechi.stimmodel}). 
We found, however, that compared to an LSTM-based EN, the linear model and the vanilla nonlinear RNN are both not
sufficiently powerful to capture the long-term dependencies in stimulation effects needed for the CPN to learn well.
Our use of an LSTM for learning an EN builds on previous work using LSTMs for predicting the evolution of local field
potentials \cite{kim.lstm} and blood-oxygen-level dependent (BOLD) hemodynamic responses \cite{guclu.lstm} over time.

\subsection{Simulation Overview}
To test the feasibility of the neural co-processor approach, we used a previously published cortical model for
grasping in a nonhuman primate (NHP) brain. Using such a model allowed us to explore some of the critical
architectural choices and training algorithms for neural co-processors, enabling us to rapidly and cheaply iterate on our
design, laying the groundwork for future \textit{in vivo} experiments. 

\begin{figure}
	\centering
	\begin{subfigure}[c]{0.69\textwidth}
		\centering
		\includegraphics[width=\textwidth]{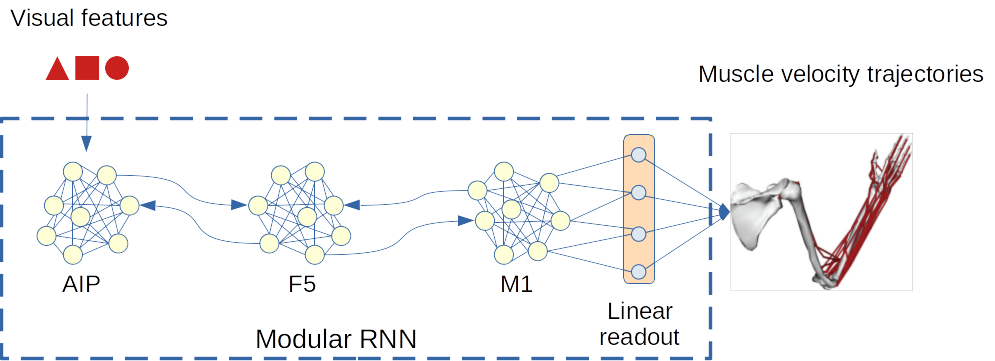}
		\caption{}
	\end{subfigure}
	\hfill
	\begin{subfigure}[c]{0.30\textwidth}
		\centering
		\includegraphics[width=\textwidth]{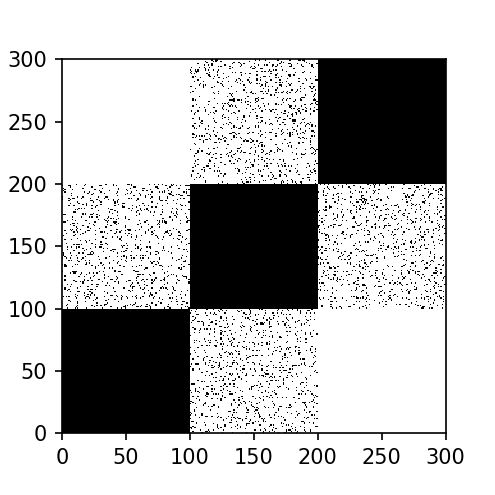}
		\caption{}
	\end{subfigure}
	\hfill
\caption{\textbf{Architecture of the Cortical Model for Grasping.} (a) Modular RNN (mRNN) used by Michaels et al.
        \cite{michaels.mrnn} to model cortical circuits involved in grasping objects.
        The emergent dynamics of the three modules correspond well to neural activity in
        primate cortical areas AIP, F5, M1, respectively, for the same task. Visual
        features of an object (derived from VGGNet) propagate forward through the
		network, conditioning the grasp on the object's size and shape. (b)
		Connectivity matrix $J$. Connected neuron pairs are indicated in black, though the
		actual network weights are floating point values. White indicates non-connections.
		Note the each module is fully connected (black squares along the diagonal) while connections to adjacent
        modules are sparse ($10\%$ of possible connections). All connections were trained using nonhuman primate
        (NHP) data to recapitulate the experimentally recorded muscle velocity trajectories for grasping different
        objects.}
\label{fig:michaels}
\end{figure}

Specifically, we use the cortical model of Michaels et al.\ \cite{michaels.mrnn}
(Fig.~\ref{fig:michaels}a). The model, which uses multiple RNNs to represent multiple interconnected cortical areas,
was trained to mimic the grasping circuits of NHP subjects engaged in a delayed reach-to-grasp task.
The model's design draws on a body of literature focused on architectures and training methods for RNNs which
seek to create artificial neural networks with activation dynamics similar to biological circuits,
including circuits for delayed grasping tasks \cite{sussillo.mrnn}. The model consists of three
``modular'' vanilla RNNs (mRNNs), representing the cortical areas AIP (anterior intraparietal cortex), F5 (ventral
premotor cortex) and M1 (primary motor cortex) respectively, and a linear readout layer from M1 producing muscle
velocities (Fig.~\ref{fig:michaels}a). Each ``module'' consists of 100 vanilla RNN neurons, with a nonlinearity
applied on the outputs. The modules are internally fully connected, and are connected to each other sparsely
(10\% connectivity in our case). The inputs to the network are visual features
representing the object to be grasped, as well as a hold signal. The visual features
intend to capture the features represented in the subject's visual cortex. They were extracted using
VGGNet \cite{simonyan.vgg} from 3D renderings of the same objects which the subjects grasped. The hold signal is
a Boolean which encodes the point in the experiment when the subject began to reach. The outputs of the
network are muscle velocities for the shoulder, arm, and hand of the subject. The actual
velocities were captured with a motion capture glove. The Michaels et al.\ network model 
was trained to recapitulate these grasping motions.\footnote{Data and trained models from this work
were supplied to us by the lead author of \cite{michaels.mrnn}. We re-implemented their model in PyTorch,
and used their trained parameters for an arbitrarily chosen subject.} 

The Michaels et al.\ cortical model implements a vision-to-grasp pipeline, from the visual
processing needed to move the hand to the appropriate position to shaping the hand for
grasping an object of a particular shape. The emergent dynamics of the model's modules, once trained,
correspond roughly to neural responses in the cortical areas AIP, F5, and M1 in a NHP subject's brain (Fig.~\ref{fig:michaels}a;
see \cite{michaels.mrnn} for details). For convenience, we will refer to the three
modules in the model using the cortical areas (AIP, F5, M1) they correspond to. For details on how this network model 
was trained, please see Supplementary Materials \ref{sup:michaelstraining}. For additional details on the task structure, see
Supplementary Materials \ref{sup:michaelstask}.

An important attribute of this cortical model is that  the simulated circuit's activity shows a relatively
clear separation for different object shapes, i.e., the  visual input is leveraged by the model to
successfully generate hand shape trajectories for grasping objects of particular shapes and sizes. As noted
below, if this visual information is prevented from propagating through the cortical modules (e.g., due to a
simulated lesion), the model can at best learn a stereotyped grasp across all object sizes and shapes. 

\subsubsection{Lesioning the cortical model causes real world failure modes}
As shown by Michaels et al. \cite{michaels.mrnn}, simulated lesions in the cortical model described
above result in error modes which resemble the effects of some natural lesions in the primate brain.
For example, a ``lesion'' involving zeroing some of the outputs of the first (AIP) module 
leads to a reaching motion generally succeeding, but finger muscle velocities show a higher
degree of error, effectively implying that the subject can reach to grasp, but cannot
tailor the grasp to the current object. This is likely due to object shape information not
being fully conveyed to the primary motor cortex (M1) module. Such a form of hand muscle spasticity
is also a common symptom in certain strokes where the subject is able to position their
hand, but is unable to form the appropriate grasp \cite{khanna.openloop, puthenveettil.hand}. 

\begin{figure}[h]
	\centering
	\begin{subfigure}[c]{0.62\textwidth}
	    \centering
	    \includegraphics[width=\textwidth]{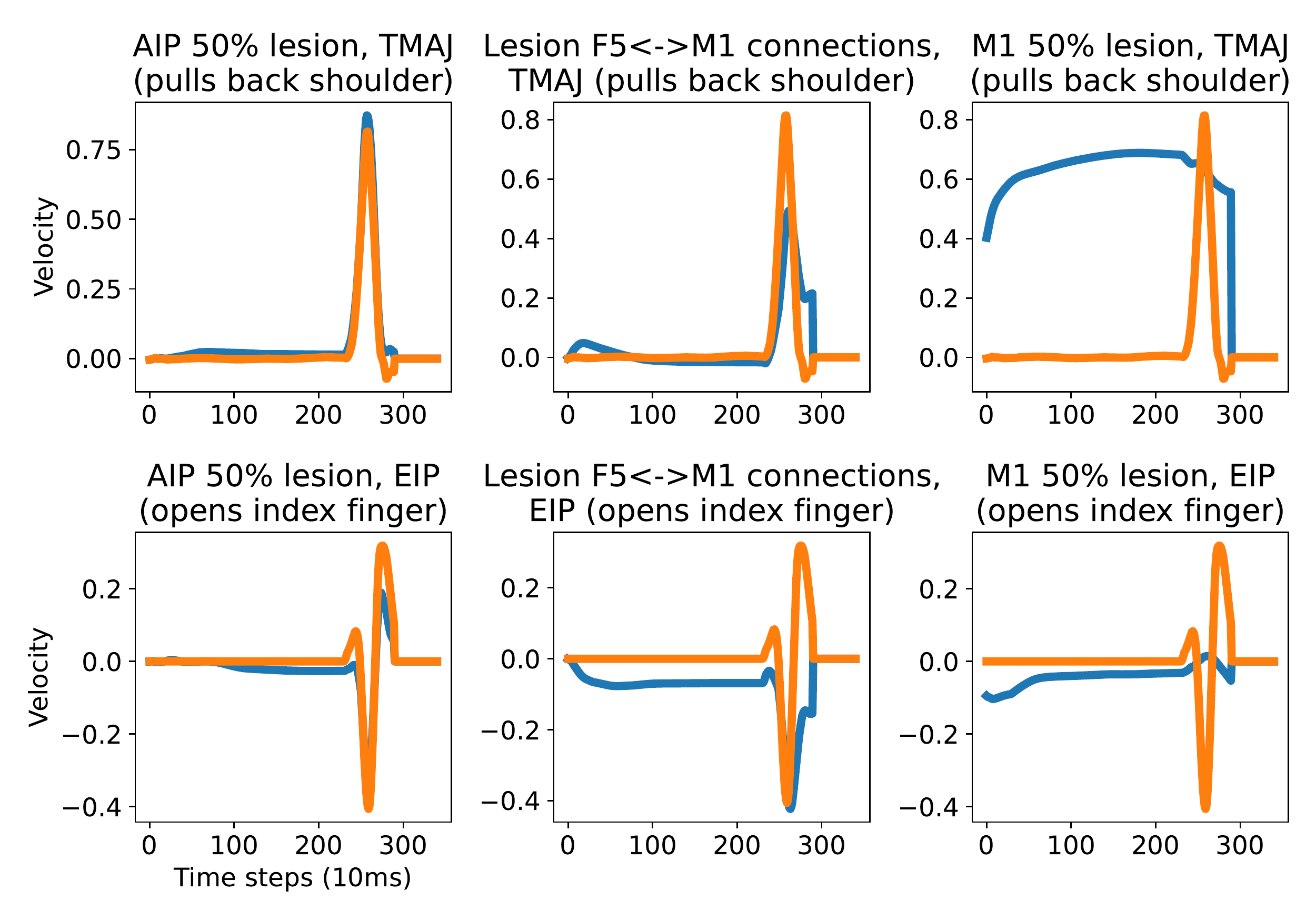}
	    \caption{}
	\end{subfigure}
	\hfill
	\begin{subfigure}[c]{0.32\textwidth}
	    \centering
	    \includegraphics[width=\textwidth]{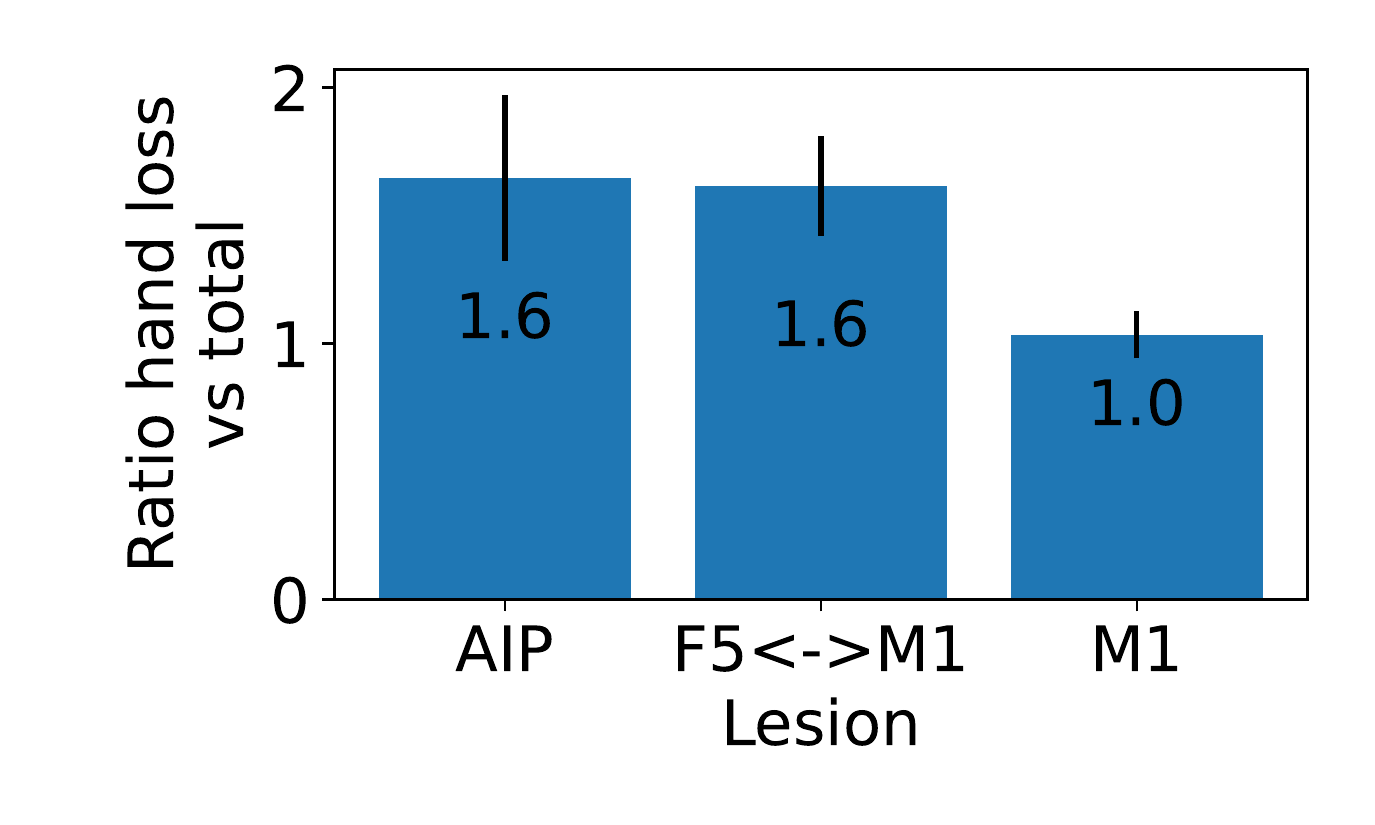}
	    \caption{}
	\end{subfigure}
	\hfill
	\caption{\textbf{Simulating strokes by lesioning different parts of the model
             lead to differential impact on grasping behavior.}
             (a) Example muscle trajectories before (orange) and after (blue)
             simulated stroke due to AIP, F5-M1 and M1 lesions for a shoulder
             muscle (top row) and a hand muscle (bottom row). Lesions which prevent
             a forward propagation of visual information tend to have a larger effect
             on hand pose (example EIP muscle shown) than on the shoulder, which is more
             involved in the reach motion (example TMAJ muscle shown). Lesions of M1
             tend to cause a significant task loss on both.
             (b) Ratio of hand muscle mean squared error (MSE) losses vs. overall MSE
             losses, for different lesion types. The loss is measured relative to the
             model's movement trajectories prior to the lesion. Average of all trials,
             with $\pm1$ stdev shown. Higher ratios for AIP and connection lesions indicate those
             lesions cause a larger loss for hand movement than overall movement, though
             their overall loss is lower than for M1 (see Table \ref{tab:results}).}
	\label{fig:lesion}
\end{figure}

On the other hand, if we lesion of a portion of the M1 module in the model,
we see a more complete loss of movement, affecting even the ability to reach for the grasp.
Finally, if we ``disconnect'' communication between the F5 and M1 modules,
we see a failure similar to an AIP lesion: the reaching movement is generally achieved, but
we see a disproportionate impact on forming the appropriate hand grasp. See Fig.
\ref{fig:lesion}a for examples of lesion impacts on velocities for individual muscles.
In Fig. \ref{fig:lesion}b we see the ratio of mean squared error (MSE) loss for hand
muscles to all muscles. Values greater than $1.0$ indicate that hand muscles are impacted more
than other muscles. Note that with the AIP and connection lesions, hand muscles show significantly
higher loss than overall loss. However, the M1 lesion shows the highest overall loss; see Table
\ref{tab:results} in Supplement materials section \ref{sup:results} for detailed lesion loss data. 

Given a particular type of lesion in the model, the goal of the co-processor is to learn
to map ``neural recordings'' (derived from current mRNN  activity in the model)
to the appropriate stimulation pattern for executing the required grasp. The
co-processor thus seeks to effectively bridge across the lesion and deliver
stimulation to enable grasping behavior tailored to the current object's shape.

In our experiments, we studied our co-processor's performance on three types of
simulated lesions:
\begin{itemize}
	\item \textbf{Loss of AIP Neurons:} We force the output of some proportion of the AIP module's
	      neurons to zero, effectively removing them from the network. This results in some amount of 
	      loss of object shape information.
	\item \textbf{Loss of F5-M1 Connections:} We prevent the propagation of
	      information between the F5 and M1 modules, effectively representing a
	      severing of the connections between the two modules. Note that the connections are sparse
	      and run in both directions, and we lesion the connections in both directions.
	\item \textbf{Loss of M1 Neurons:} We force the output of some proportion of the M1 module's
	      neurons to zero. Here, the lesion may make it impossible for the co-processor
	      to find a perfect solution since the loss of M1 neurons may make it impossible to
	      activate muscles in the same ways as before the lesion. However, in the Results section,
        we show that some recovery is still possible.
\end{itemize}

\subsubsection{Simulated network exhibits long running and stable dynamics}
To compute stimulation patterns that optimize for a task, the co-processor must
learn to adapt to the dynamics of the network it is stimulating. Biological neural
networks as well as our simulated grasping network exhibit long range changes in dynamics
due to stimulation. A perturbation of the network (i.e., due to stimulation) will cause
changes in neuronal activations long after the stimulation has been applied, sometimes far
from the site of stimulation. Our cortical grasping network model exhibits the same
behavior.

To illustrate this, suppose we apply a small, one-time perturbation
to the hidden states of 10 randomly chosen neurons in the output (M1) module at some point
in time during a trial. By repeating this experiment many times, we can understand what the
distribution of long-running effects tends to look like on the output muscle velocities.

\begin{figure}[h]
	\centering
	\begin{subfigure}[c]{0.48\textwidth}
	    \centering
	    \includegraphics[width=\textwidth]{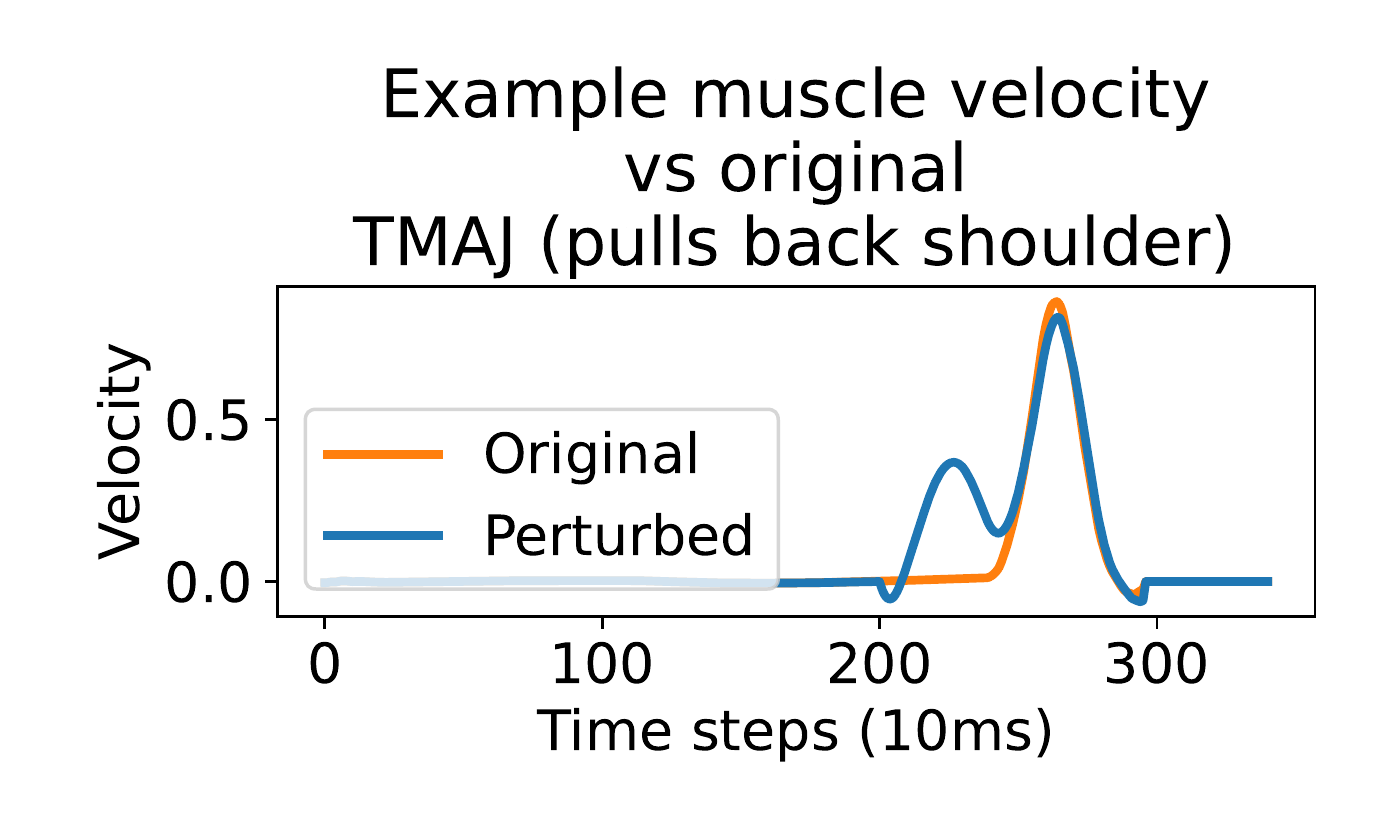}
	    \caption{}
	\end{subfigure}
	\hfill
	\begin{subfigure}[c]{0.48\textwidth}
	    \centering
	    \includegraphics[width=\textwidth]{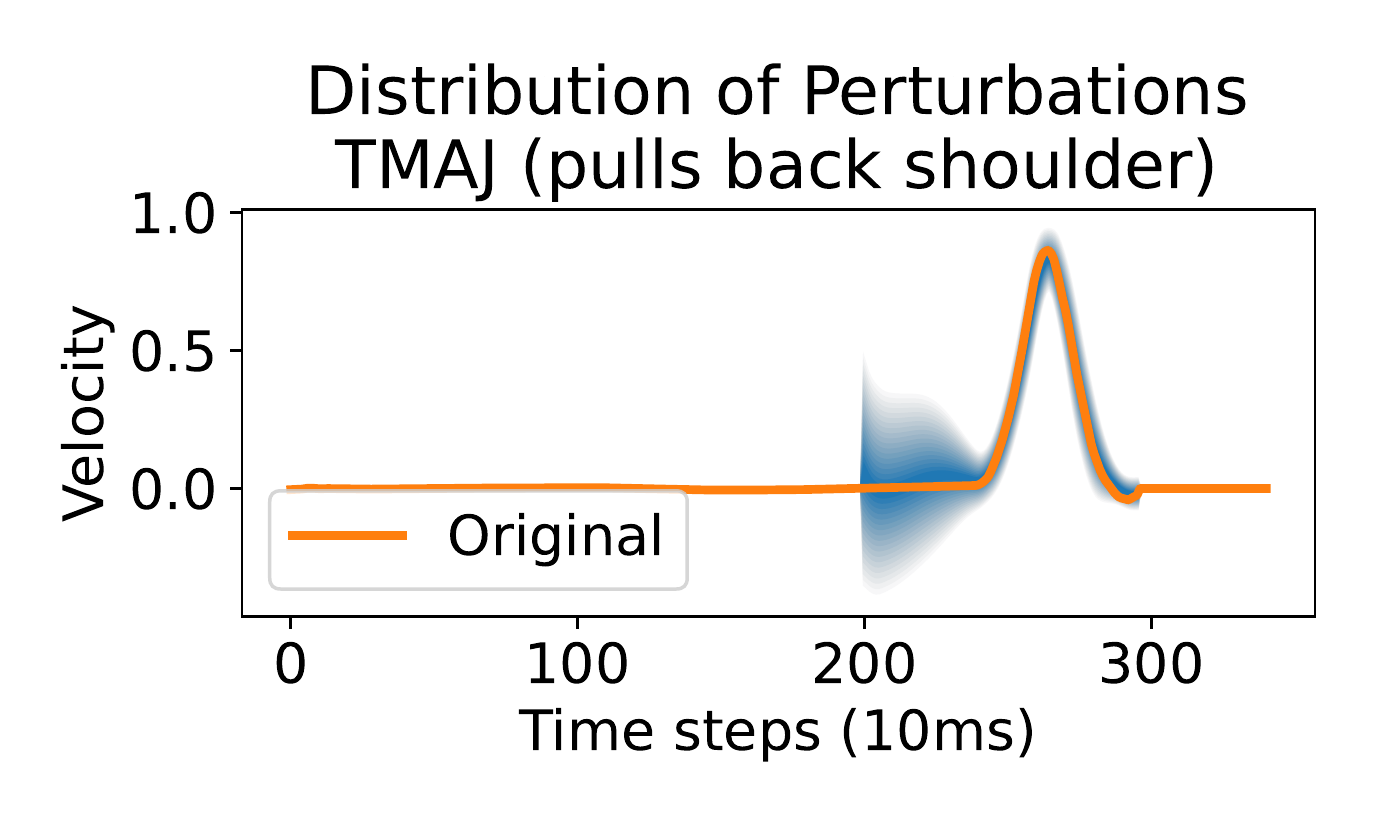}
	    \caption{}
	\end{subfigure}
	\hfill
	\begin{subfigure}[c]{0.48\textwidth}
	    \centering
	    \includegraphics[width=\textwidth]{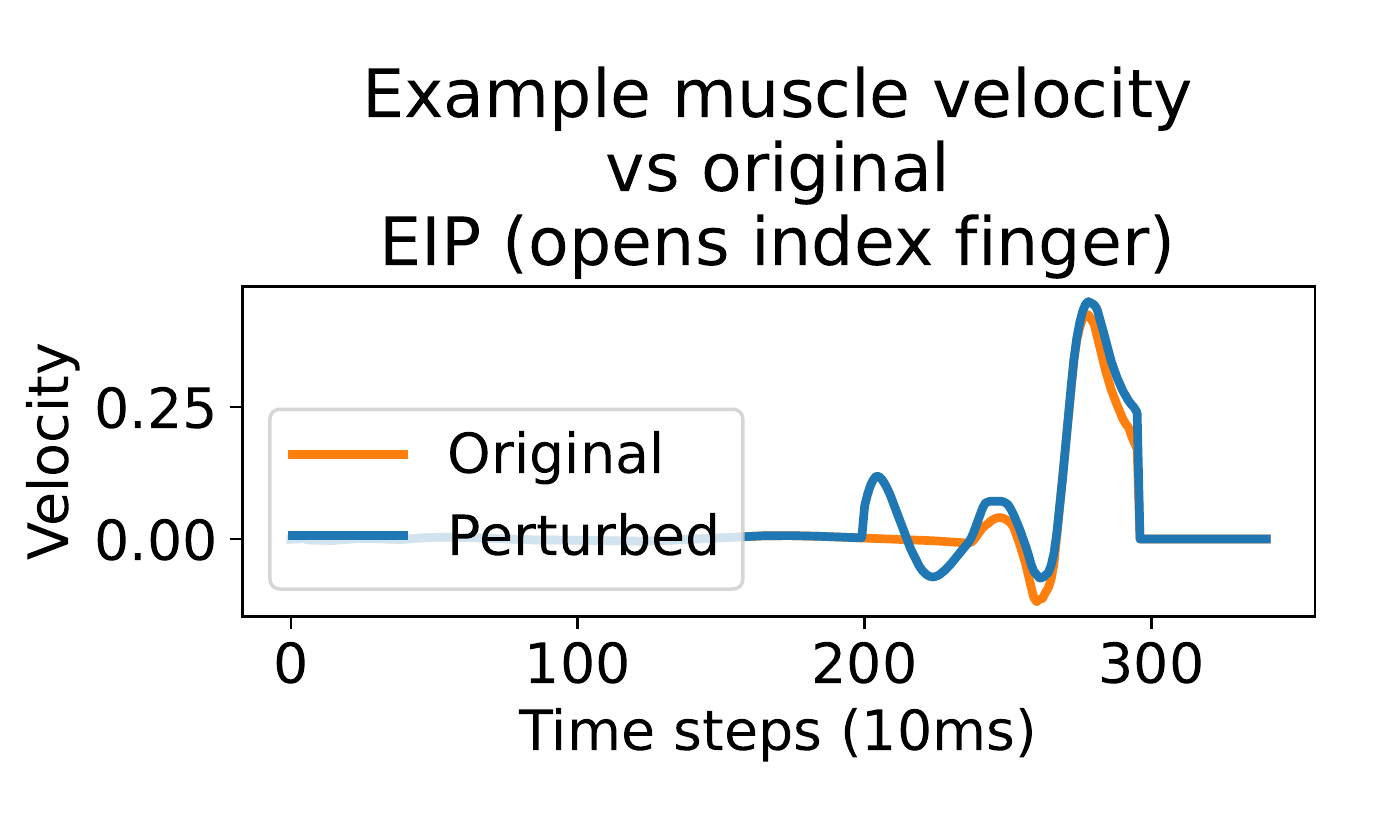}
	    \caption{}
	\end{subfigure}
	\hfill
	\begin{subfigure}[c]{0.48\textwidth}
	    \centering
	    \includegraphics[width=\textwidth]{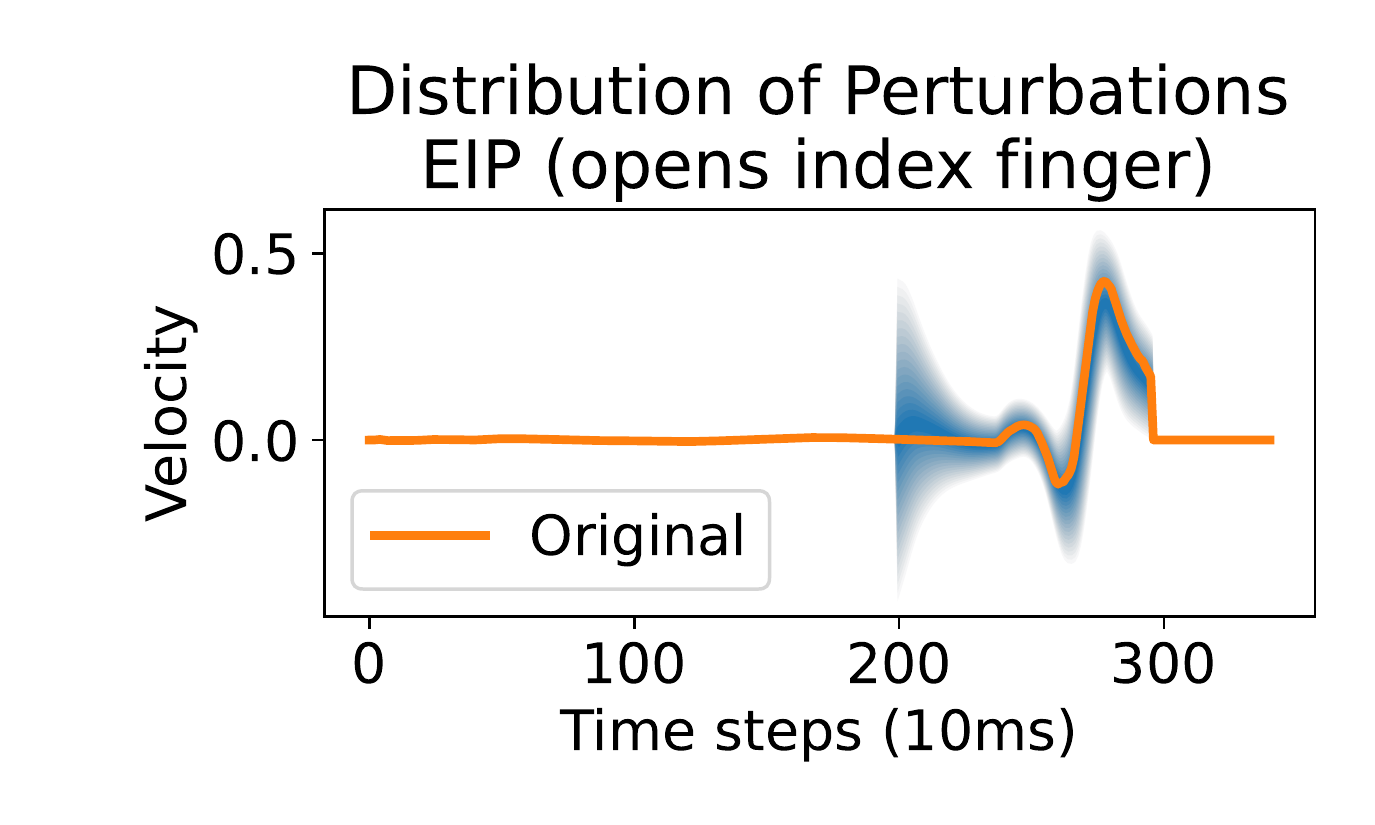}
	    \caption{}
	\end{subfigure}
	\hfill
	\caption{\textbf{Perturbations of the simulated cortical network (i.e. due to stimulation)
	         have long-running effects on muscle activation.} Our results are consistent with
	         the fixed point analysis of the network in \cite{michaels.mrnn}, showing that the network
	         exhibits stability. It tends towards the original trajectory after perturbation.
	         Instantaneous random perturbation of a random group of 10 M1
	         neurons performed at time t=2s. (a, c) Single trial example for TMAJ, EIP muscles
	         respectively. (b, d) Distribution of effects across n=1000 samples on TMAJ, EIP muscles
	         respectively. Probability distribution shown to +/- 2 stdevs. Network exhibits similar
	         behavior when perturbance is applied at other time steps.}
	\label{fig:dynamics}
\end{figure}

In Fig.~\ref{fig:dynamics} we see that even a single, one-time perturbation in
the network has effects dozens of time steps later. Our co-processor will need to learn to take
these dynamics into account. As we will show in the next subsection, the problem our co-processor
faces in our simulations is in fact even harder than this since we also incorporate
a model of stimulation effects on the neural circuit which includes both spatial and temporal
smoothing.

\subsubsection{Stimulation model}
In our experiments, we stimulate neurons only in the output area (here M1) since the co-processor's
purpose is to improve external task performance, which it is able to do with
appropriate stimulation of the M1 output region of the network model. In a more general setting, it is
conceivable that a co-processor could stimulate other areas of the brain as well to improve task
performance downstream, or to probe the brain to better reveal the user's intent (i.e. object
shape), but we leave these directions to future work.

To make our stimulation model more realistic, rather than allowing stimulation to directly change the
output of single neurons, we simulated how stimulation may affect a network of neurons using a
model that incorporates aspects of both spatial and temporal smoothing. Our intent was not to create a
biophysical model of stimulation (our model does not arrange neurons in a volume to allow for such detailed
simulation). Instead, the stimulation model approximates the effects of \textit{in vivo} biological
stimulation (e.g., extracellular electrical stimulation) by diffusing the effects of stimulation across
many neurons and across time. Our EN must thus learn to approximate this stimulation model,
in addition to approximating the cortical dynamics of the brain and the mapping of those dynamics onto
the grasping outputs.\footnote{In Supplementary Materials section \ref{sup:passthrough}, we explore  
other stimulation models.}

Our stimulation model uses a function which receives as input the stimulation parameter
vector $\theta$ from the CPN. The function performs temporal smoothing using a simple exponential decay model by
adding its current input to an exponentially decaying sum of inputs that decays towards $0.0$ at some
rate (see Equation ~\ref{eq:temporalstim} below). The effect of stimulation is thus not instantaneous,
but rather decays with time. Likewise, to simulate the difficulty of stimulating only single neurons,
we applied Gaussian smoothing to map $\theta$ (16-D in our experiments) to changes in the activations of
a large number of the simulated M1 neurons (100 neurons in our experiments). If we assume each element
of the $\theta$ represents the stimulation parameter for a single electrode, our Gaussian smoothing
operation emulates how stimulation may affect the neurons in its vicinity more than it affects neurons
further away. To accomplish this, we assume our model neurons are aligned along a spatial dimension
arbitrarily and fix $\sigma=1.75$.\\

\noindent The resulting equations which define our stimulation model are:

\begin{equation}
\alpha_{t} = \tau\alpha_{t-1} + \theta_{t-1} \label{eq:temporalstim}
\end{equation}
\begin{equation}
s_{t} = C\alpha_{t}
\end{equation}

\begin{itemize}
	\item $\alpha$: the 16 dimensional internal state 
        of the stimulation function
	\item $\tau$: our decay rate for temporal smoothing, which we set arbitrarily to $0.7$
	\item $C$: a fixed $100 \times 16$ Gaussian smoothing matrix containing a single 1-D Gaussian
	           in each column, to implement spatial smoothing and spread of stimulation to the
	           stimulated neural population. 
	\item $s_{t}$: the spatiotemporally smoothed stimulation vector whose elements denote how
	               much stimulation is applied to each neuron at time step $t$ (see
	               Equation~\ref{eq:neural-activation} below).
\end{itemize}

\noindent The governing equations of an mRNN with stimulation then become:

\begin{equation}
x_{t+1} = Jx_{t} + Iv_{t} + s_{t} + b
\label{eq:neural-activation}
\end{equation}
\begin{equation}
a_{t} = tanh(x_{t})
\end{equation}
\begin{equation}
y_{t} = La_{t} + l
\end{equation}

\begin{itemize}
	\item $x$: the hidden state of each model neuron
	\item $J$: recurrent weight matrix
	\item $I$: input weight matrix
    \item $v_{t}$: input 
	\item $b$: activation bias
	\item $L$, $l$: parameters of the linear readout layer
	\item $y_{t}$: the output of the network
\end{itemize}

\begin{figure}[h]
	\begin{subfigure}[c]{0.45\textwidth}
		\centering
		\includegraphics[width=\textwidth]{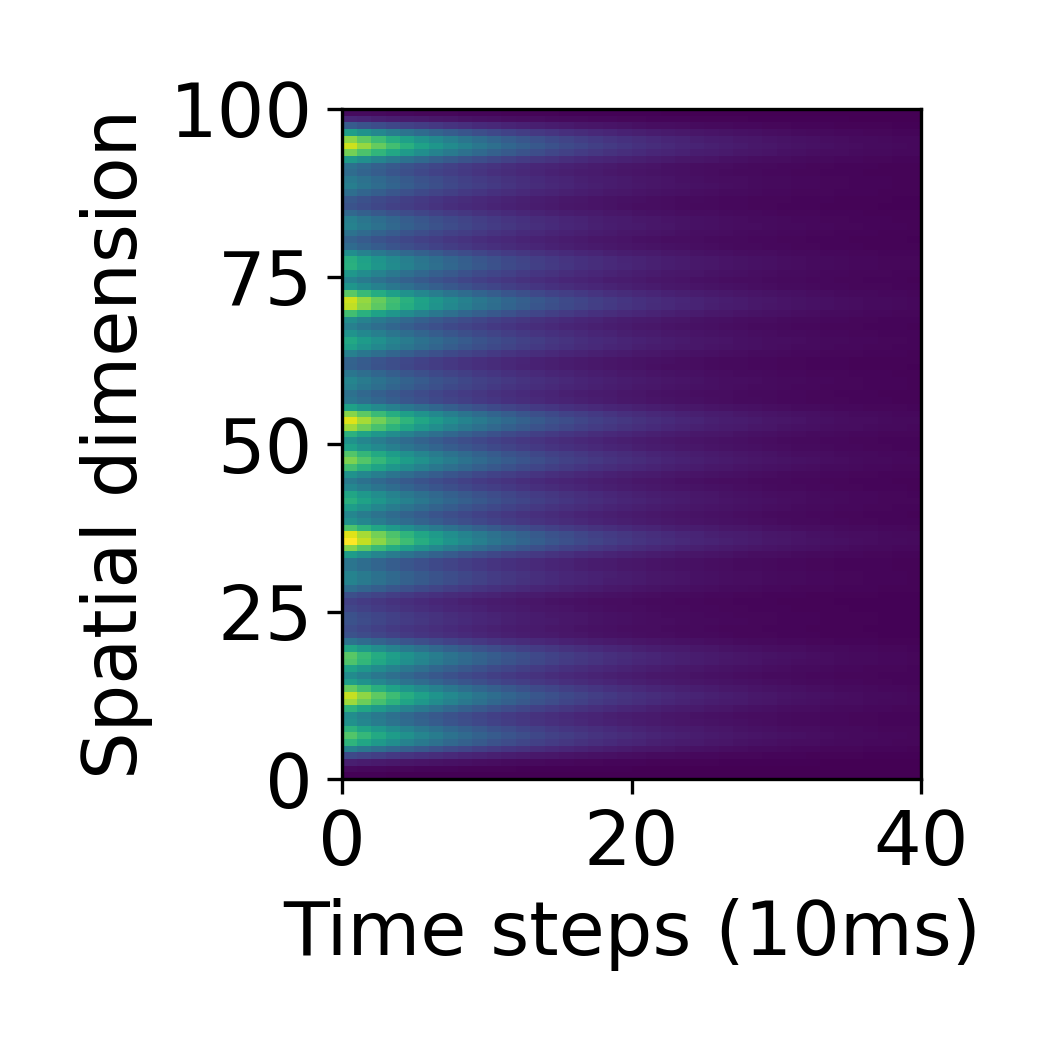}
		\caption{$\sigma=1.875$}
	\end{subfigure}
	\hfill
	\begin{subfigure}[c]{0.45\textwidth}
		\centering
		\includegraphics[width=\textwidth]{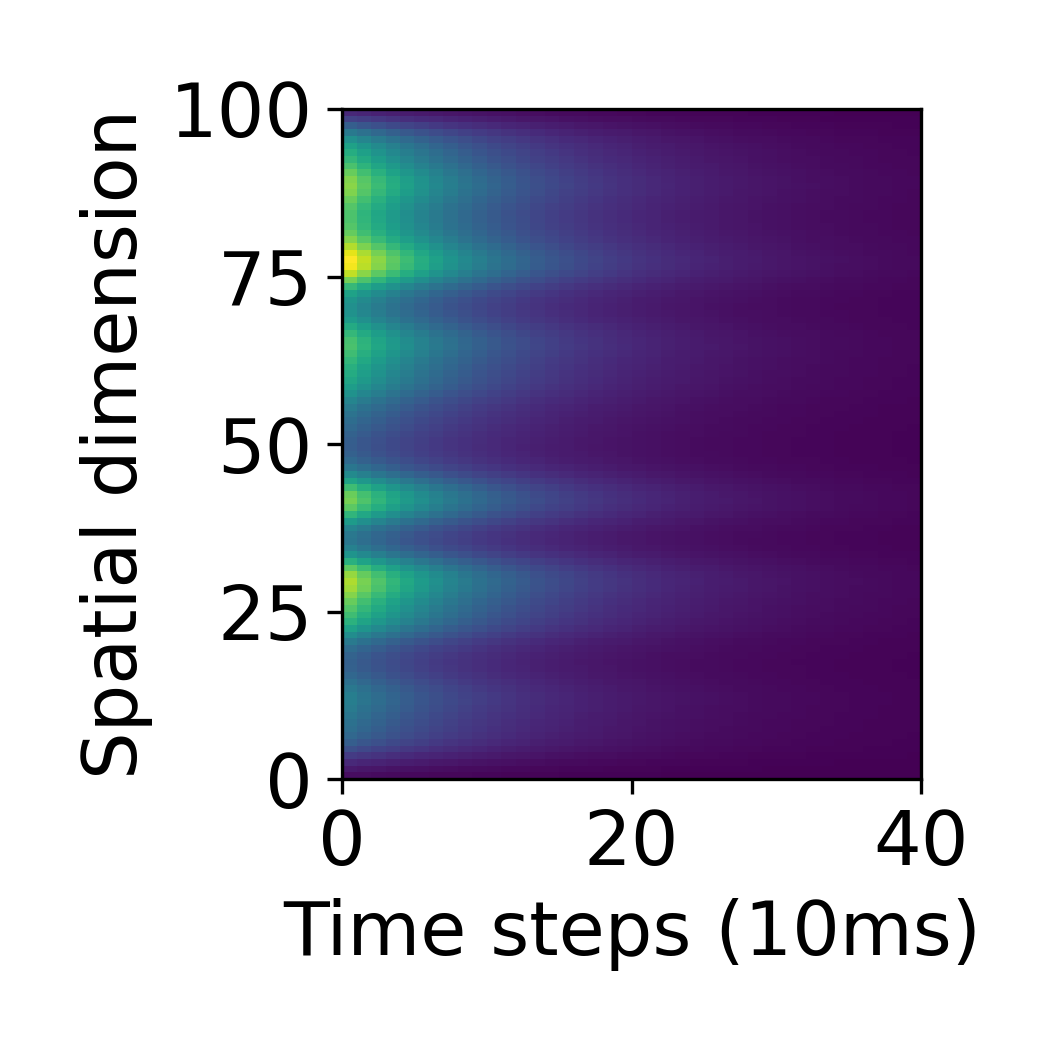}
		\caption{$\sigma=2.875$}
	\end{subfigure}

	\caption{\textbf{Simulating the effects of stimulation on a recurrent network across space and time.}
          The stimulation model performs spatial and temporal smoothing of a 16-dimensional
	         vector of stimulation parameters $\theta$ onto 100 neurons in a recurrent network.
	         The 16 elements of the vector represent in-effect 16 electrodes located
	         evenly along a spatial dimension, along which the 100 neurons are arranged
	         arbitrarily. We show here the effects of two randomized $\theta$ with two respective smoothing
	         parameters $\sigma$ (a. $\sigma=1.875$, b. $\sigma=2.875$). After $t=0$, $\theta$ is the zero
		   vector. Color values indicate the magnitude of the value summed into each neuron's hidden
		   state at that time step.}
	\label{fig:stim_single}
\end{figure}

\begin{figure}[h]
	\centering
	\includegraphics[width=\textwidth]{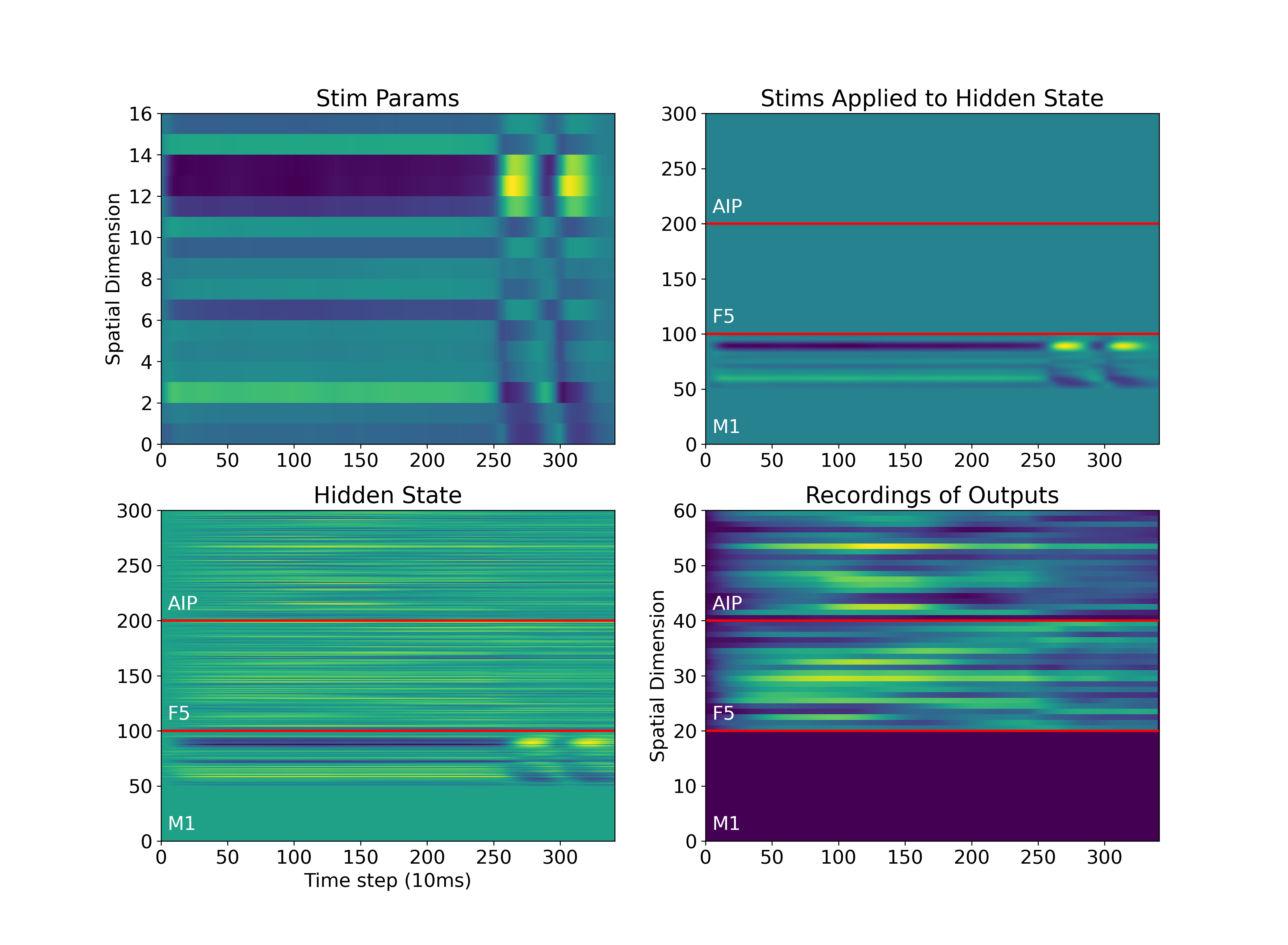}
	\caption{\textbf{Example of stimulation and recording for a single trial.} Here, M1
	has been lesioned 50\%, indicated by hidden states of M1 neurons with a value of zero (bottom
	left panel, lowest part of the plot). We record neural activities (using the recording model) from only
	 the AIP and F5 modules, as explained in Section \ref{sec:experiments}. Stimulation is applied
	to M1 to drive the network's output. Stimulation is mostly constant until approximately t=250,
	when the hold signal was lifted and the reach began.}
	\label{fig:stim_and_obs}
\end{figure}

\noindent Fig. \ref{fig:stim_single} depicts an example where
$\theta$ is non-zero at $t=0$ and zero for all other times, illustrating the effects of 
stimulation on the network across space and time.  Fig. \ref{fig:stim_and_obs} illustrates the
effects of stimulation during a stroke simulation. In this case, the trial is one where $50\%$
of M1 neurons have been  lesioned (their outputs are zero), the inputs to the CPN are from the
AIP and F5 modules, and stimulation is applied to the M1 module. 

\subsubsection{Recording model}
To simulate recordings of our model neurons, we used a recording model that assumed that the model neurons
are laid out along a single spatial dimension with a given number of electrodes spread evenly apart.
We then applied Gaussian convolution, where the Gaussian kernel is centered at each electrode position.
Thus, the recording from each ``electrode'' is a Gaussian-weighted average of the activities of all model neurons
in the given module. Fig.~\ref{fig:stim_and_obs} (bottom right panel) shows an example of recordings obtained
using this recording model; note that the Gaussian averaging provides a potentially ambiguous view of the
activities of the underlying neurons due to the inherently destructive nature of averaging, making it
more challenging for the co-processor to interpret the neural activity and produce an appropriate
stimulation pattern. Our simulated recordings are similar in spirit, though not actually modeling,
local field potentials (LFPs) recorded extracellularly in biological neural tissue. In Supplementary Materials section
\ref{sup:observability}, we show that as we vary the observability of our ``simulated brain'' via this recording
model, the results remain largely the same, up to the point where the objects to be grasped can no longer be
distinguished from the neural recordings due to excessive averaging. 

\subsubsection{Simulating co-adaptation by the brain}
To demonstrate the co-processor's ability to co-adapt with the brain as it adapts to the co-processor's
stimulation, we modify the cortical grasping network model's parameters (synaptic weights and biases) throughout
the co-processor's training. We use the standard error backpropagation algorithm to adapt the grasping network to
minimize task loss for the object grasping task (using $PyTorch$'s implementation of the $AdamW$ optimizer). With each
trial, task loss is calculated based on the NHP training data and backpropagated through the mRNNs as they receive
stimulation. The learning rate was set arbitrarily to a relatively low rate of $1\mathrm{e}{-7}$ so that the network
adapts more slowly than the co-processor.

\subsubsection{Simulating recovery prior to co-processor use}
After a stroke, the human brain has the ability to learn and recover to some extent the behaviors
affected by the stroke. We simulated this ability in our grasping network model by re-training the 
network for the grasping task after lesioning it. For our simulated lesions which zero-out 
the outputs of neurons, we found that the cortical model has sufficient redundancy built into it that lesioning it by
inactivating large numbers of neurons often leaves enough remaining degrees of freedom that a
nearly full recovery can occur. We therefore explored lesions of  the model (specifically lesions of AIP and M1 modules)
under two conditions: (a) No co-adaptation: the cortical model was not re-trained after the lesion, allowing us to study
how much of lost function the co-processor can learn to restore on its own, and (b) Co-adaptation: the cortical model
was re-trained after the lesion, allowing us to study how the co-processor copes with non-stationarity as our simulated
brain recovers from its lesion.

In the case of a lesion that prevents communication between the F5 and M1 modules, information about
the input object's shape cannot propagate forward in the network to allow shaping of the hand for
grasping. However, the cortical model can learn a stereotyped grasp during the recovery period. After this recovery
period, a co-processor can help further boost grasp accuracy by forward-propagating object shape information
from AIP and F5 to M1, acting as an artificial neural bridge. We explore this application of the
co-processor in one of our experiments below.

\subsubsection{Simulating a non-stationary recording function, e.g., sensor drift}
Over time, implanted sensors may drift from true readings. We explored the co-processor's ability to adapt to
non-stationarity in the recording function. We performed an experiment where the recording function
includes an added bias term which changes over time, according to a random process. Each epoch, we
add a random value to each element of the bias term, drawn from a zero-mean Gaussian distribution, causing the
``recordings'' to drift over time. See Supplementary Materials section \ref{sup:drift} for further details.

\subsection{Training Paradigm}
\label{sec:training}

To train the CPN to generate appropriate stimulation patterns for grasping a given input object of a
particular shape, we would like to use a learning algorithm such as the backpropagation algorithm and
minimize the errors between the current generated grasp (in terms of muscle velocities) and the desired target
grasp. However, backpropagation requires the error to be in terms of CPN output stimulation error. We cannot  
backpropagate grasp error through biological networks in the brain to get the error needed to train the CPN. 
This motivates our use of the EN as a model for the transformation from stimulation to muscle velocities:
we backpropagate grasp error through the EN, while keeping its parameters fixed, and use this
backpropagated error to change the parameters of the CPN in order to minimize grasp error.

\begin{figure}
	\centering
	\begin{subfigure}[c]{0.48\textwidth}
		\centering
		\includegraphics[width=\textwidth]{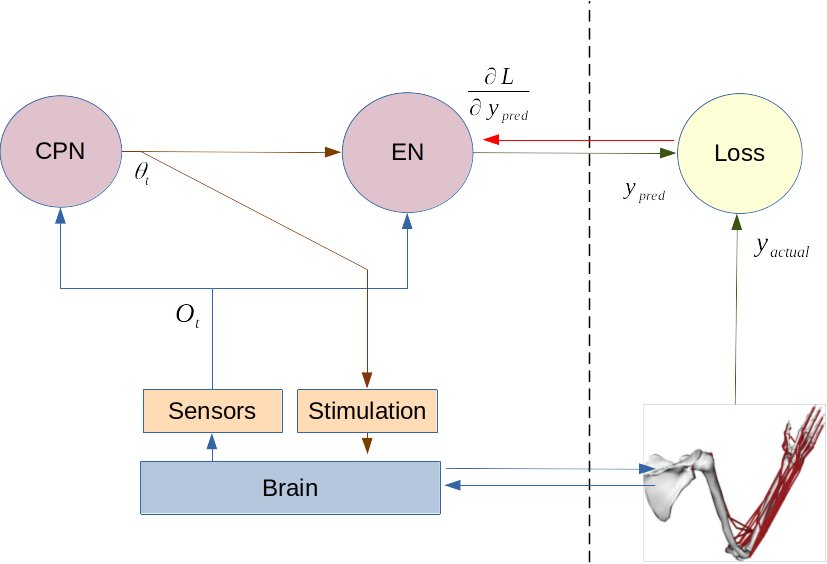}
		\caption{}
	\end{subfigure}
	\hfill
	\begin{subfigure}[c]{0.48\textwidth}
		\centering
		\includegraphics[width=\textwidth]{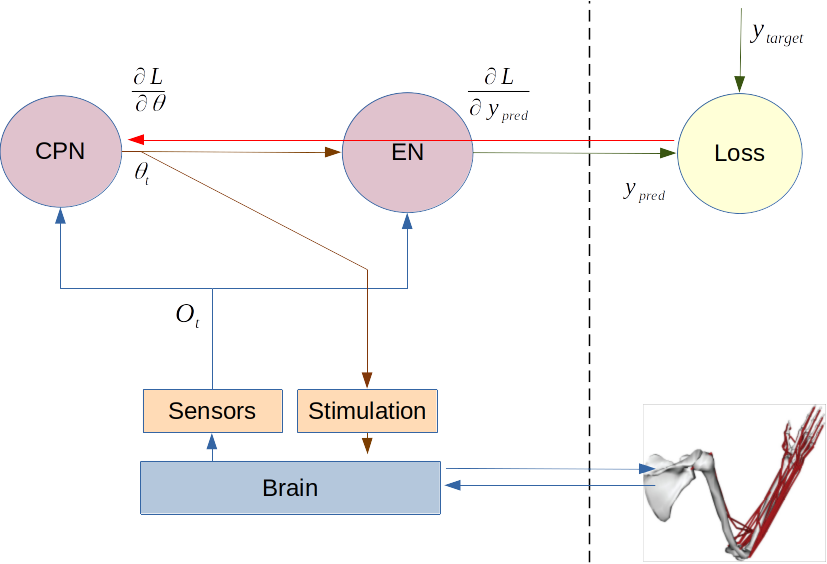}
		\caption{}
	\end{subfigure}
	\hfill
\caption{\textbf{Training the neural co-processor.} (a) EN training phase: 
         backpropagate the error between actual vs predicted muscle velocities through the EN to update the EN's parameters.
         (b) CPN training phase: backpropagate the error between EN's predicted output vs target muscle velocities, via the
         EN (without changing its parameters) and through the CPN, to update the CPN's parameters. Note that we are effectively
         treating the EN's predicted output as the actual muscle velocity output, and therefore it is important to have a good
         pre-trained EN.}
\label{fig:training}
\end{figure}

Using the EN to train the CPN requires careful interleaving of EN and CPN training epochs. EN training
epochs concentrate on updating the EN, based on observations of the effect of stimulation on the cortical
network model's output. The loss function when training the EN is the mean squared error (MSE) loss between
the EN's prediction and the  actual output of the cortical model (i.e., muscle velocities). During the CPN
training epochs, we keep the EN's parameters fixed and use the EN to train the CPN. Specifically, we
backpropagate the MSE loss between the EN's predicted output, and the desired output (target muscle velocities
for grasping the input object) through the CPN. Fig. \ref{fig:training} illustrates the EN and CPN training
processes.

\subsubsection{Training and testing data sets}
For training and evaluation of our model, we use the same data as in Michaels et al.
\cite{michaels.mrnn}. The data consist of:
\begin{itemize}
    \item The input visual features and hold signal. We input these to the simulated
          cortical model;
    \item The object identity. We use this to calculate how well the co-processor
          can differentiate objects; see Results section below;
    \item The muscle velocity data for the trial, as extracted from the data glove of a nohhuman primate 
          during the NHP trial, and as processed by Michaels et al.
\cite{michaels.mrnn}. This is our target output.
\end{itemize}

For each training session, we hold out a random sample of $20\%$ of the data to act as our validation data set; we
use the other $80\%$ for training.

\subsubsection{EN training}
\label{sec:en_training}

For the EN to be useful for training the CPN, it must accurately predict the behavioral effects of stimulation
produced by the CPN. In addition, backpropagating through the EN must yield gradients which are useful for
changing the weights of the CPN in order to produce better stimulation patterns for minimizing the error.
We discovered that this latter property does not occur simply by virtue of the former. Specifically, an EN
can be trained to high levels of predictive power, even for random stimulation, and to orders of magnitude
lower loss than the task loss, while at the same time,  backpropagation through the EN still yields gradients
which lead to unstable training of the CPN. 

We hypothesize that the crux of the EN training problem is one of over-fitting: the EN may be trained
to achieve high predictive power on a data set of stimulation examples, but due to 
the high dimensionality of our stimulation parameters $\theta$, and the dynamics of the
network being stimulated, the function learned by the EN may not provide gradients suitable
for the stimulation inputs generated by the CPN during its training.

To address this problem, we first structure the training dataset for each EN training epoch
in a specific way. In each epoch, we include:
\begin{itemize}
	\item training examples using stimulation inputs generated by the current CPN;
	\item training examples using stimulation inputs generated by a small collection of CPNs obtained
	      by adding zero-mean Gaussian noise to the current CPN's parameters; and
	\item training examples with white noise stimulation inputs.
\end{itemize}

Such a training paradigm is designed to cover a sufficient variety of examples to prevent overfitting,
and to do so in a way that explores the neighborhood of the current CPN's parameter space. Augmenting
a CPN-generated data set with white noise examples alone was not sufficient to stabilize CPN training,
but including additional examples generated by CPNs in the neighborhood of the current CPN helped train
the EN to perform well in that neighborhood and produce useful gradients for the CPN. ``Neighborhood''
here refers to the region of CPN parameter space near the current CPN. This may cause the EN to be
overfit, i.e., fit to the local area of CPN parameter space but exhibiting poor predictive power in
other areas of the space. However, we solve this problem by retraining the EN, interleaved with CPN
training. We show the effect of this paradigm in our results below
(section~\ref{sec:results_con}).

We composed each batch of EN training data as follows: \\

\begin{tabular}{|l|l|}
\hline
\textbf{Source}                                                                                                                            & \textbf{Proportion of dataset} \\ \hline
Data from current CPN                                                                                                       & 10\%                           \\ \hline
\begin{tabular}[c]{@{}l@{}}Data from current CPN with parameter noise\end{tabular} & 60\%                           \\ \hline
Data from white noise stimulation                                                                                                        & 30\%                           \\ \hline
\end{tabular} \\

Additionally, we use decoupled weight decay regularization \cite{loshchilov2017decoupled} to
mitigate overfitting, and a carefully chosen learning rate schedule. The schedule begins with
a higher learning rate initially ($4\mathrm{e}{-3}$), ramping down to a lower rate ($1\mathrm{e}{-4}$).
The learning schedule is based on the most recent prediction error on the validation portion of the
data set. We found that using rates much lower or higher than these rates in any training phase
may cause the EN learning to not converge. EN training proceeds until the network's prediction error
is lower than a threshold, defined as a fraction of the current CPN's task loss. Section
\ref{sec:interleaving} provides further details on when we switch between training the EN and CPN.

\subsubsection{CPN training}
Once an EN is properly trained, we can use the EN as a surrogate for the cortical model's transformation
of stimulation patterns and neural activity to grasping behavior (i.e., muscle velocities). We use a large number
of randomly sampled trials from the original Michaels et al.\ task \cite{michaels.mrnn} and generate
stimulation patterns using the current CPN alone. These stimulation patterns are passed through the EN
to generate predictions of output muscle velocities, which are compared to the desired target velocities
to compute the error. We backpropagate this error through the EN (without modifying its parameters) to
generate training gradients for the CPN. Fig.~\ref{fig:training} illustrates this training process.

We found that the CPN appears to train in two phases. In the first phase, it is largely learning the structure
of the reach-to-grasp task, e.g., that the muscles need to remain at a position during the hold period until the
reach begins. During this early training phase, the learning rate can be quite high ($1\mathrm{e}{-3}$). Later, the CPN
begins to learn the mapping between object shape information and the stimulation patterns needed to approximate
the target grasp for that object. This later phase of training requires a learning rate 2-3 orders of magnitude
lower than in the first phase ($1\mathrm{e}{-6}$ to $5\mathrm{e}{-5}$). See Section~\ref{sec:results} for more details.

\subsubsection{Interleaved CPN/EN training, and adapting to non-stationarity}
\label{sec:interleaving}
Having defined the training procedures for the EN and CPN, we can define a training
protocol combining those two. We train the two in alternation, creating
a new EN each time we enter an EN training phase. We explored the possibility
of reusing an existing EN but found that retraining an EN was no more fast than training a new one.
Training a new EN also allows us to adapt to the brain's non-stationarity, which requires us to
determine when the current EN is no longer suitable for training the current CPN. 
Our algorithm retires the current EN and trains a new one when one of two conditions
is satisfied: (1) the EN's prediction error is sufficiently above some fraction of the CPN's task loss;
or (2) if CPN loss does not improve (or gets worse) over a sufficient number of recent training steps 
(this is a common stopping criterion in machine learning). Additional details can be found in the
Supplementary Materials section ~\ref{sup:encpninter}.

\subsection{Experiments}
\label{sec:experiments}

We performed eight experiments to investigate the co-processor's ability to learn under a variety of
conditions. We studied three types of lesions to the cortical network model (simulated stroke) and for
each lesion type, we turned on or off (a) brain/co-processor co-adaptation (brain recovery during
co-processor use), (b) brain recovery {\em before} co-processor use, and (c) non-stationarity due to
sensor drift during co-processor use. The eight experiments are summarized in the following table 
(a blank entry under a condition (e.g., Co-adaptation?) means ``No'' while an X means ``Yes''): \\

\begin{table}[h]
\centering
\begin{tabular}{|l|l|c|c|c|}
\hline
 & \textbf{Lesion}                                           & \multicolumn{1}{l|}{\textbf{\begin{tabular}[c]{@{}l@{}}Co-adaptation?\end{tabular}}} & \multicolumn{1}{l|}{\textbf{\begin{tabular}[c]{@{}l@{}}Prior recovery?\end{tabular}}} & \multicolumn{1}{l|}{\textbf{\begin{tabular}[c]{@{}l@{}}Sensor\\ drift?\end{tabular}}}\\ \hline
1              & 50\% AIP loss                                             &                                                                                                 &                                                                                                &                                                                                                 \\ \hline
2              & 50\% AIP loss                                             & X                                                                                               &                                                                                                &                                                                                                 \\ \hline
3              & 50\% M1 loss                                              &                                                                                                 &                                                                                                &                                                                                                 \\ \hline
4              & 50\% M1 loss                                              & X                                                                                               &                                                                                                &                                                                                                 \\ \hline
5              & 100\% connection loss F5\textless{}-\textgreater{}M1 &                                                                                                 &                                                                                                &                                                                                                 \\ \hline
6              & 100\% connection loss F5\textless{}-\textgreater{}M1 & X                                                                                               &                                                                                                &                                                                                                 \\ \hline
7              & 100\% connection loss F5\textless{}-\textgreater{}M1 & X                                                                                               & X                                                                                              &                                                                                                 \\ \hline
8              & 100\% connection loss F5\textless{}-\textgreater{}M1 & X                                                                                               &                                                                                                & X                                                                                               \\ \hline
\end{tabular}
\end{table}

Each of the above experiments used the same set of input and output data as was used by
Michaels et al. \cite{michaels.mrnn} to train the cortical model that we use. The dataset
contains a total of 502 trials, sampled uniformly across 42 object classes. In each experiment,
we held out a random sample of 20\% of the dataset for validation.

\subsection{Stopping criteria}
In these experiments, we train the co-processor until one of the following two stopping criteria is
satisfied (in Supplementary Materials section \ref{sup:stopping_criteria}, we probe how these criteria
affect the results). The two criteria are:

\begin{itemize}
	\item The percent change in average task loss between two consecutive ranges of 500 epochs
	      is below a threshold, indicating minimal benefits of further training. The choice of
        threshold varies by experiment, due to the significant differences in setup. See Supplementary
        Material section \ref{sup:stopping_criteria} for further details.
	\item The training run has exceeded some number of epochs. This also varied by experiment.
\end{itemize}

\section{Results}
\label{sec:results}

For each lesioning experiment, we track two metrics to characterize the learning progress. First, we
track the task loss, which is an MSE loss that measures the co-processor's ability to restore
movement towards the target trajectory. We compute the loss based on the error between the muscle
velocity trajectory output by the lesioned cortical network model and the ground truth muscle velocity
data from the Michaels et al. experiment \cite{michaels.mrnn}. In our results, we characterize this loss
in terms of percent loss recovery, i.e., difference between lesioned and healthy performance.
Specifically the percent loss recovery metric is defined as:

\begin{equation}
  \mbox{percent loss recovery} = (1 - \frac{L - L_{healthy}}{L_{lesioned} - L_{healthy}}) \times 100 
\end{equation}

\noindent where $L$ is the current task loss, and $L_{healthy}$ and $L_{lesioned}$ are the task losses before the
lesion was applied and after the lesion was applied (but before any recovery) respectively. Thus 0\% represents
percent loss equal to the lesioned brain (prior to any recovery) while 100\% represents percent loss equal to the
healthy brain. Note that it is possible for this metric to  become negative if current task loss $L$ exceeds
$L_{lesioned}$ (e.g., if the stimulation is not beneficial and makes task performance worse than the
performance with the lesioned brain).

Second, we measure the degree to which the lesioned cortical model, when coupled with the
co-processor, successfully differentiates the grasps for different object classes. We compare the ratio of
total variation to within-class variation of muscle velocities, for grasps executed by the lesioned and
the healthy cortical network. We define the grasp separability metric $S$ as:

\begin{equation}
	S = \frac{\sigma_{a}}{\sigma_{w}} - \frac{\sigma_{a,h}}{\sigma_{w,h}}
\end{equation}

\noindent where $\sigma$ measures the mean standard deviation, across time, and across all data points in
a given sample, $\sigma_{a}$ is the standard deviation of a dataset, and $\sigma_{w}$
is the average of the within-class standard deviations ($h$ indicates healthy). An $S$ value of 0.0
indicates that the grasps for the different object classes vary to the same degree as the healthy network.
A negative value indicates the grasping motions are more similar between the object classes; likewise a
positive value indicates the grasps are more dissimilar.

The grasp separability metric $S$ allows us to differentiate between overall task performance improvement
and the ability of the co-processor to condition the grasp on the input object based on visual information.
To successfully grasp in the real world, the hand must be preformed appropriately for
the shape of the object being grasped, and must close around it appropriately.
Note that it is possible for the structure of the delayed reach-to-grasp task to be potentially learned
by the co-processor, (e.g., to hold muscle velocities to 0.0 for the initial part of each trial) without
learning to also differentiate the various object shapes. $S$ provides a useful metric in this regard for
tracking changes in grasp separability as opposed to only overall changes in task loss.

\subsection{Experiment 1: AIP 50\% lesion}
\begin{figure}[h]
\centering
\includegraphics[scale=1]{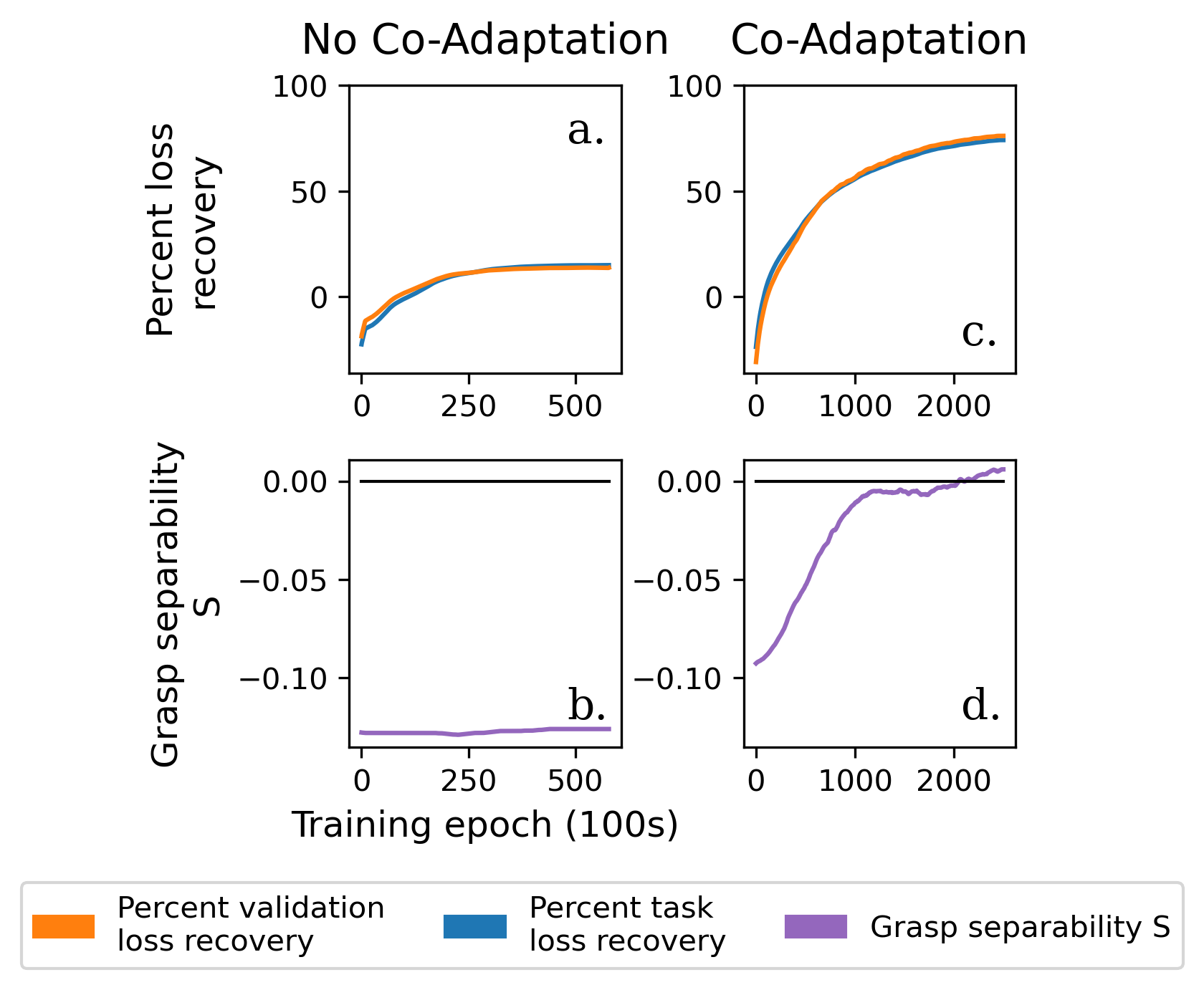}
\caption{Co-Processor Performance Results: AIP 50\% Lesion. a. Percent loss recovery for the non-coadaptive
         experiment for the task and validation datasets. 0\% represents task loss equal to the loss for the
         lesioned cortical model, prior to any recovery. 100\% represents loss equal to that for the healthy
         cortical model. Negative initial values are due to the newly-initialized CPN causing
         worse performance than the lesioned model prior to recovery. b. Grasp separability metric ($S$) for the non-coadaptive
         experiment. Negative $S$ values indicate grasps are more similar between object types while 
         positive values indicate they are more distinct. c., d. show the performance of the neural co-processor for the
         co-adaptive experiment.}
\label{fig:results_aip}
\end{figure}

Our first experiment tested the ability of the co-processor to compensate for the loss of a significant
fraction (50\%) of the model neurons in the AIP module of the cortical model; this module is responsible for encoding
object shapes from visual inputs. Loss of AIP model neurons results in the cortical model, without further adaptation,
being unable to condition its grasp on the object shape, after largely succeeding in reaching toward the object.\footnote{
In Supplementary Materials section \ref{sup:results}, lesioned loss is lower in this experiment than in others: the subject
continues to reach successfully, but does not form the hand properly.} Likewise, the co-processor cannot observe sufficient
object shape encoding information due to the AIP lesion. As a result, as shown in Fig. \ref{fig:results_aip}a, significant
recovery is not possible. Fig. \ref{fig:results_aip}b shows that the co-processor never learns to separate object classes
($S$ value never approaches or exceeds $0.0$).

In the co-adaptive case, the brain and co-processor together find a solution that reaches $76\%$ recovery towards healthy
performance. This demonstrates that the co-processor can adapt to the non-stationarity
of the mapping between the muscle outputs and the CPN-delivered stimulation.
As the brain adapts, it learns to encode the input visual information, which in turn allows the co-processor
to observe that information through its ``recordings.'' The co-processor can then leverage that information to condition
stimulation, improving task performance (Fig. \ref{fig:results_aip}c). Note, though, that this simulation is not designed
to indicate if the co-processor speeds up recovery, or provides better results than natural recovery, since we aren't
modeling the timeline of natural recovery. With this lesion design, simulated recovery would in-fact allow near-complete
recovery of task performance. As a result, this experiment is designed  only to demonstrate co-adaptation.

In Fig.~\ref{fig:results_aip}d, which plots the grasp  separability metric $S$, we see that initially
the co-processor and the simulated brain do not strongly separate the object classes, resulting in negative $S$ values,
but as the co-processor  training proceeds, the metric exceeds 0.0, indicating separability. This can be attributed to
the co-processor and the simulated brain learning how to map the input visual information about object shape to the
appropriate grasp for
the object.

In Figs.\ \ref{fig:results_aip}a and c, the initial percent loss recovery values are negative: this is because the
newly-initialized CPN in the co-processor delivers stimulation that initially causes task performance to be worse
than the performance of the lesioned simulated brain; this is corrected with further training of the CPN. 

\subsection{Experiment 2: M1 50\% lesion}
\begin{figure}[h]
\centering
\includegraphics[scale=1]{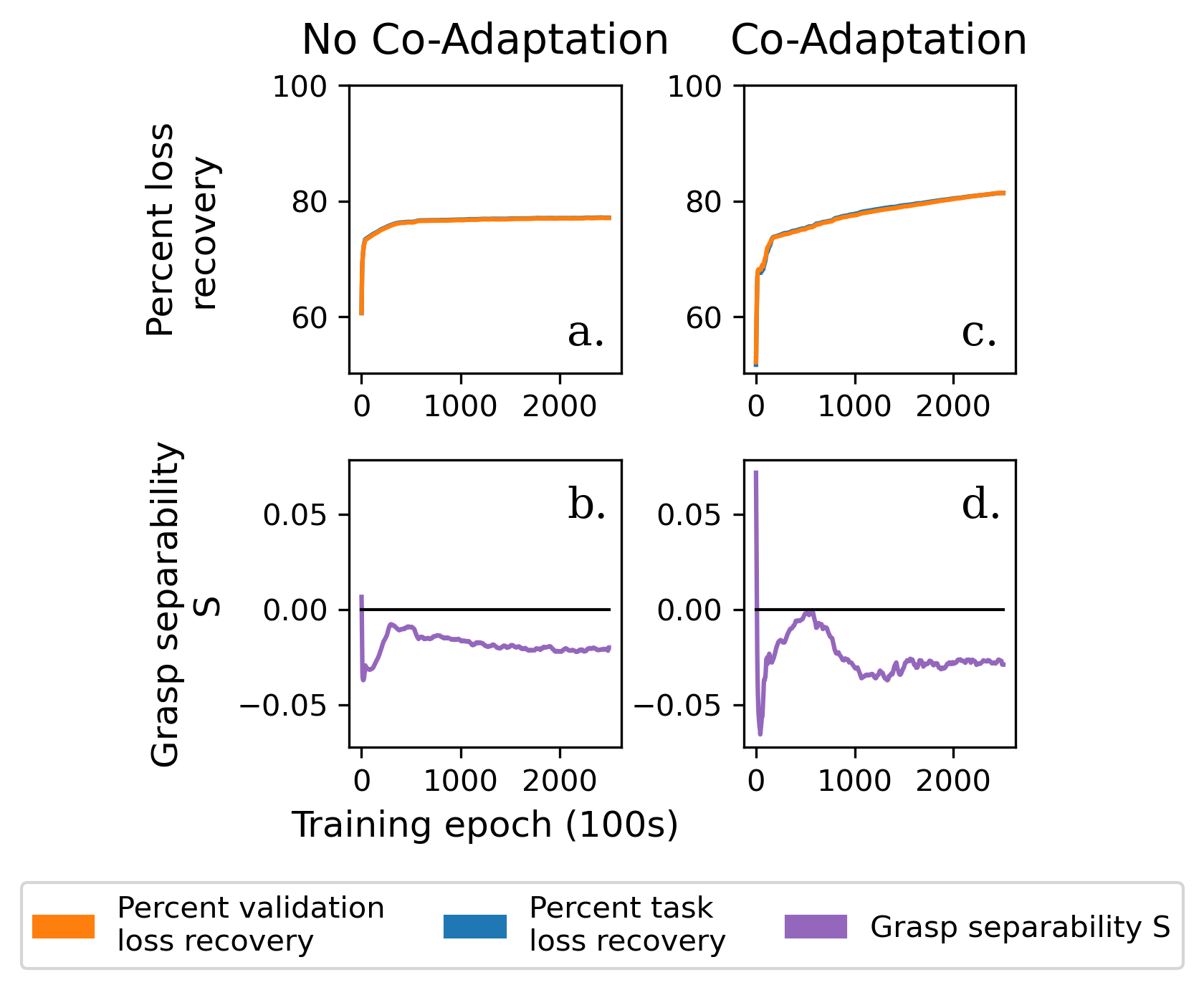}
\caption{Co-Processor Performance Results: M1 50\% Lesion. a. Percent loss recovery
         for the non-coadaptive experiment for the task and validation datasets. See caption for Fig.~\ref{fig:results_aip}
         for interpretation of positive and negative values. b. Grasp separability metric (S) for the non-coadaptive experiment.
         See caption for Fig.~\ref{fig:results_aip} for interpretation of positive and negative values. c., d. show the
         performance of the neural co-processor for the co-adaptive experiment. 
         }
\label{fig:results_m1}
\end{figure}

In this experiment, we lesioned 50\% of the M1 module of the cortical model.
As shown Fig.\ \ref{fig:results_m1}a, in the non-coadaptive case, training quickly plateaus at 75\%, and does not improve
thereafter. This suggests the lesion inactivated some of the degrees of freedom needed to
control the output layer, resulting in a reduced ability to recreate the different target grasps
for different objects (Fig.\ \ref{fig:results_m1}b). 


In the co-adaptive case (Fig. \ref{fig:results_m1}c), performance likewise hits an inflexion point at 75\%, but
continues to improve slowly. The ability to improve further is tied to the
``natural'' recovery occurring in the co-adapting cortical model, which restores some of the
degrees of freedom needed for improving grasping performance (Fig.\ \ref{fig:results_m1}d). The co-processor 
n this case adapts successfully to this non-stationarity being caused by the recovery process. 

\subsection{Experiment 3: F5 and M1 connection lesion}
\label{sec:results_con}

\begin{figure}[h]
\centering
\includegraphics[scale=1]{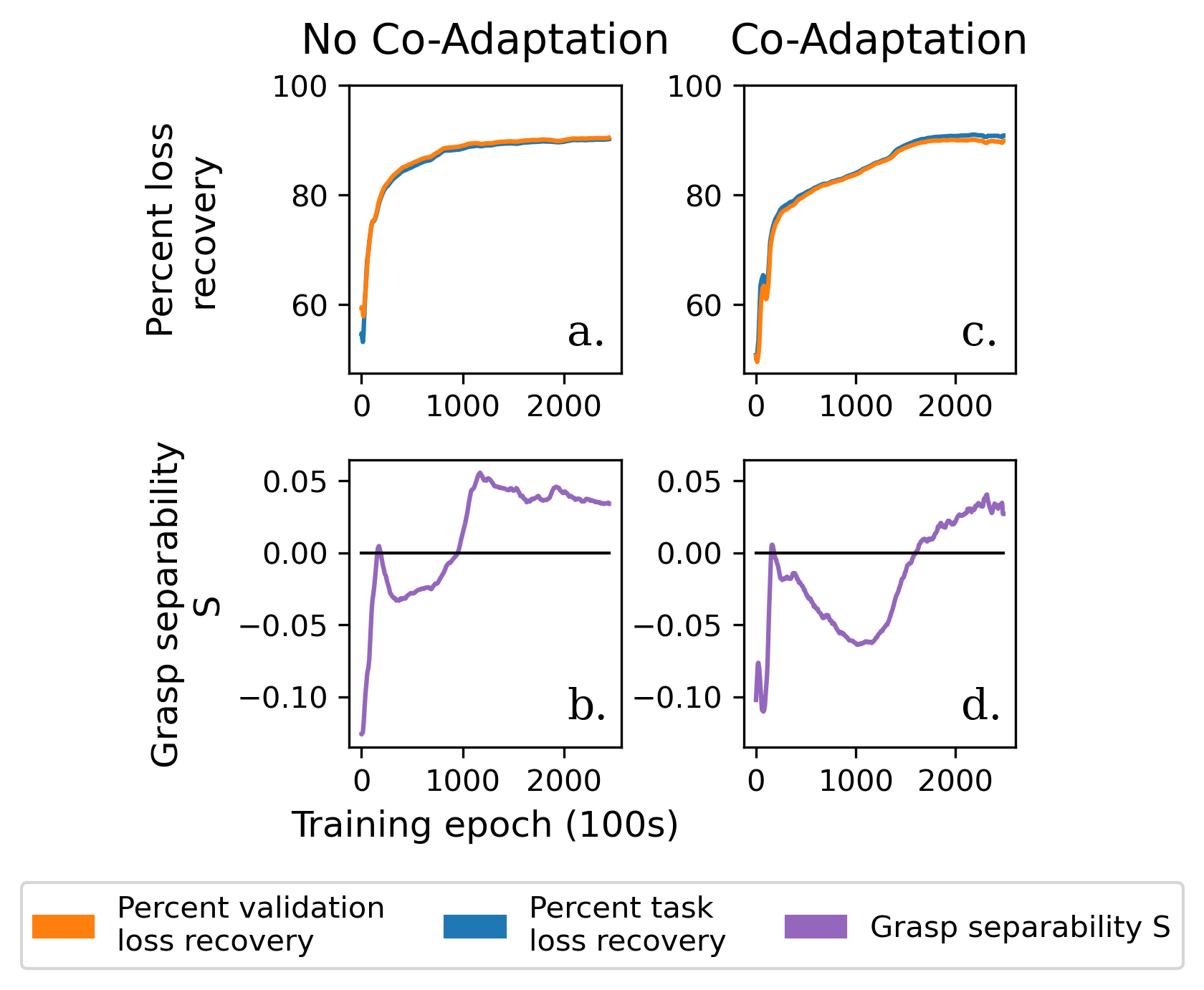}
\caption{Co-Processor Performance Results: F5-M1 Connection Lesion. 
          a. Percent loss recovery for the non-coadaptive experiment for the task and validation datasets.
          See caption for Fig.~\ref{fig:results_aip} for interpretation of positive and negative values.
          b. Grasp separability metric (S) for the non-coadaptive experiment. See caption for Fig.~\ref{fig:results_aip} 
          or interpretation of positive and negative values. c., d. show the performance of the neural
          co-processor for the co-adaptive experiment.
         }
\label{fig:results_con}
\end{figure}

\begin{figure}[h]
\centering
\includegraphics[scale=1]{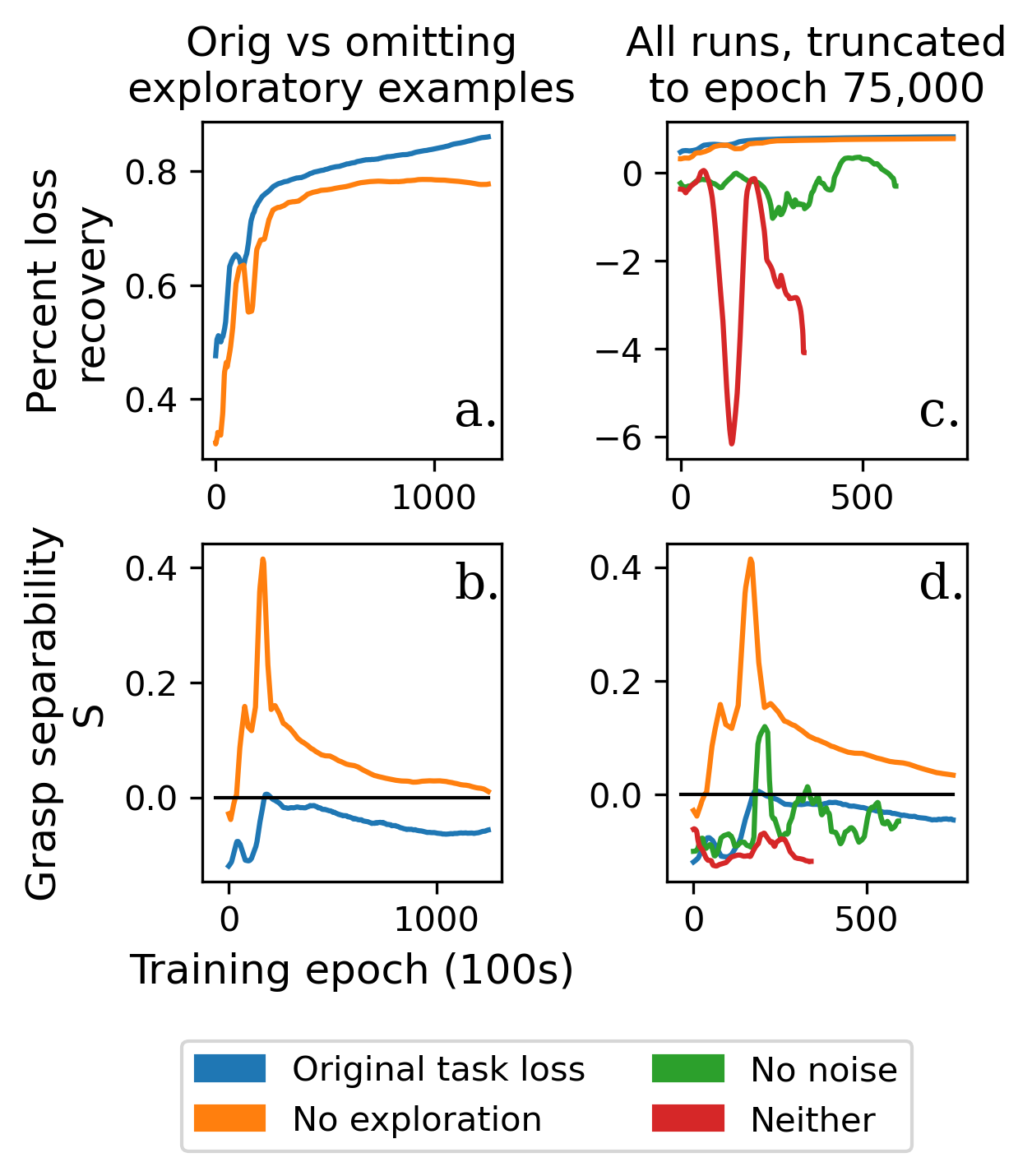}
\caption{Co-Processor Performance Results: F5-M1 Connection Lesion: Use of exploratory
         and randomized examples in the training data set lead to stabilized
         and faster training. a., b. Without exploratory examples in the
         neighborhood of the CPN, we have unstable and slower training. In
         this case, the data set is composed of $10\%$ examples drawn from the
         CPN, and $90\%$ randomized examples (``No exploration'').
         c., d. Using only examples from the CPN (``Neither''; no exploratory or randomized examples) or $10\%$ CPN
         and $90\%$ exploratory examples (``No noise''; no randomized examples) also
         leads to unstable training. In c.\ and d., we truncate the experiments and use different
         axes to allow for clearer plots, but show the data from a.\ and b.\ for
         comparison.
         }
\label{fig:results_con_no_explore}
\end{figure}

In this experiment, we disconnected F5 from M1 completely to test whether the co-processor can
act as a bridge between the two areas to appropriately convey information required for the grasping
task from one area to the other. As in the other experiments, the co-processor's task performance in this
experiment improved quickly (Fig.~\ref{fig:results_con}a and c), but it took much longer to refine the
stimulation patterns to enable object differentiation (Fig.~\ref{fig:results_con}b and d). 

Since feedback from the output cannot ``backpropagate'' to the F5 and AIP modules, co-adaptation and
learning in this experiment only affects the M1 module. As a result, the co-processor's performance in
this experiment demonstrates that its learning algorithm is capable of adapting to non-stationarity
in the cortical model's mapping between stimulation parameters and the behavioral output (muscle velocities
for grasping). These results contrast  with the results of Experiment 1, where the non-stationarity affected
the mapping between the visual inputs and the AIP outputs.

As seen in Fig.~\ref{fig:results_con}b and d, grasp separation for objects is initially low and unstable,
as the co-processor begins to learn the task, but later stabilizes as the co-processor begins to
leverage observed brain activity to differentiate object shapes, gradually meeting or exceeding the
pre-lesion grasping performance.

To illustrate the effect of the EN training dataset composition as described in section~\ref{sec:training}, we
performed additional experiments on the F5-M1 connection lesion where we varied the dataset composition. As
outlined in section~\ref{sec:training}), the training data set for the EN is composed of stimulation examples
drawn from the current CPN, examples drawn from other CPNs with parameters in the neighborhood of the current
CPN in parameter space (referred to as ``exploratory examples''), and randomized stimulation. In
Fig.~\ref{fig:results_con_no_explore}, we see the effect of removing the latter two types of examples.
Exploratory examples proved necessary to stabilize training, and to achieve higher levels of recovery. Removing
the examples of randomized stimulation resulted in even more unstable training, where no improvement in task
performance occurs.

\subsection{Experiment 4: Connection lesion with recovery}
It is often the case that after a stroke, the brain recovers some of the function lost immediately after the
stroke. We simulated this natural recovery in our cortical  network model by lesioning the F5-M1 connections
and then training the lesioned network on the grasping task. Complete recovery is impossible because the M1 module's
connections to F5 (both forward and backward) are disconnected. The cortical network model in this case learns
an object-agnostic stereotypical grasp. We tested whether a co-processor can subsequently improve grasping performance
beyond natural recovery by learning to convey to M1 via stimulation the object-related information from AIP and F5
necessary to tailor the grasp to the current input object's shape.

\begin{figure}[h]
\centering
\includegraphics[scale=1]{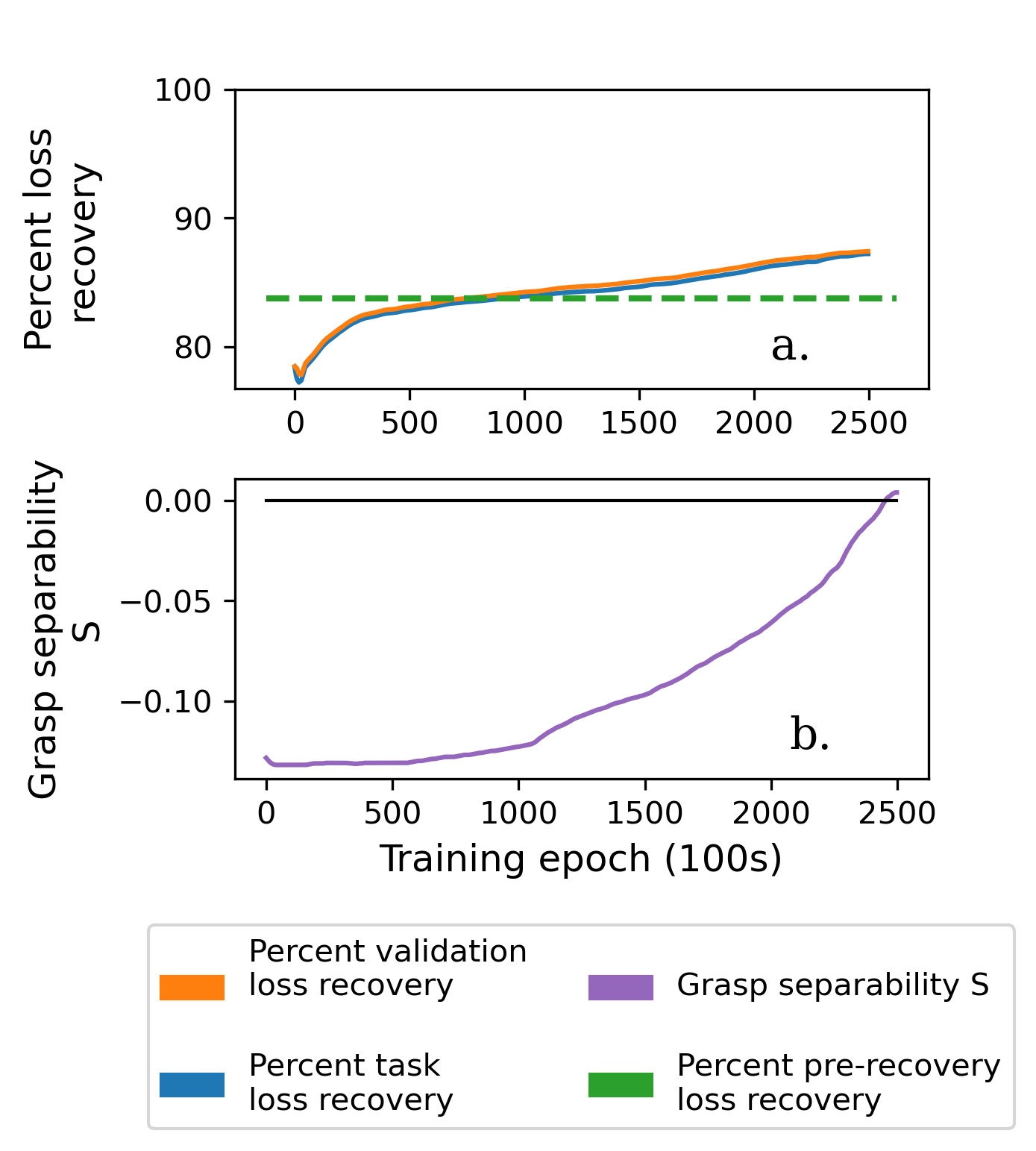}
\caption{Co-Processor Performance Results: Connection Lesion and Natural Recovery before Co-Processor Use.
         a. Percent loss recovery for the task and validation datasets. The horizontal dashed line is the task
         performance after both lesioning and natural recovery but before co-processor use. Initially the
         co-processor's stimulation causes task performance to be worse than the performance after natural
         recovery, but eventually the co-processor's performance exceeds this level. b. Grasp separability
         metric (S). Negative S values indicate grasps are more similar between object types while positive values
        indicate they are more distinct.
        }
\label{fig:results_recov}
\end{figure}

We found that the co-processor can successfully improve task performance beyond the natural
recovery after the simulated ``stroke'' (Fig.~\ref{fig:results_recov}a). The dashed line
indicates task loss after initial recovery. As in some experiments above, task performance is
initially worse than after recovery, due to stimulation by an  initially untrained co-processor. However
the co-processor gradually learns to drive the task loss lower, as it learns to forward-propagate information
from the ``earlier'' parts (AIP and F5) of the simulated brain. Also, note that the grasp separation metric $S$
returns to a healthy value towards the end of co-processor training
(Fig.~\ref{fig:results_recov}b).

\subsection{Experiment 5: Connection lesion with sensor drift}
Sensor-related non-stationarities are
often seen in real-world neural recording systems due to a variety of factors, from impedance
changes due to sensor movement to scar tissue formation. In our final experiment, we again lesioned the F5-M1
connections and allowed co-adaptation by the simulated cortical model but additionally, we allowed the sensor
readings to drift over time. This drift was modeled in the recording function, where the readings had a non-stationary
zero point across epochs. We added a bias term which we updated between epochs according to a random process
(see Supplementary Materials section~\ref{sup:drift} for details). 

\begin{figure}[h]
\centering
\includegraphics[scale=1]{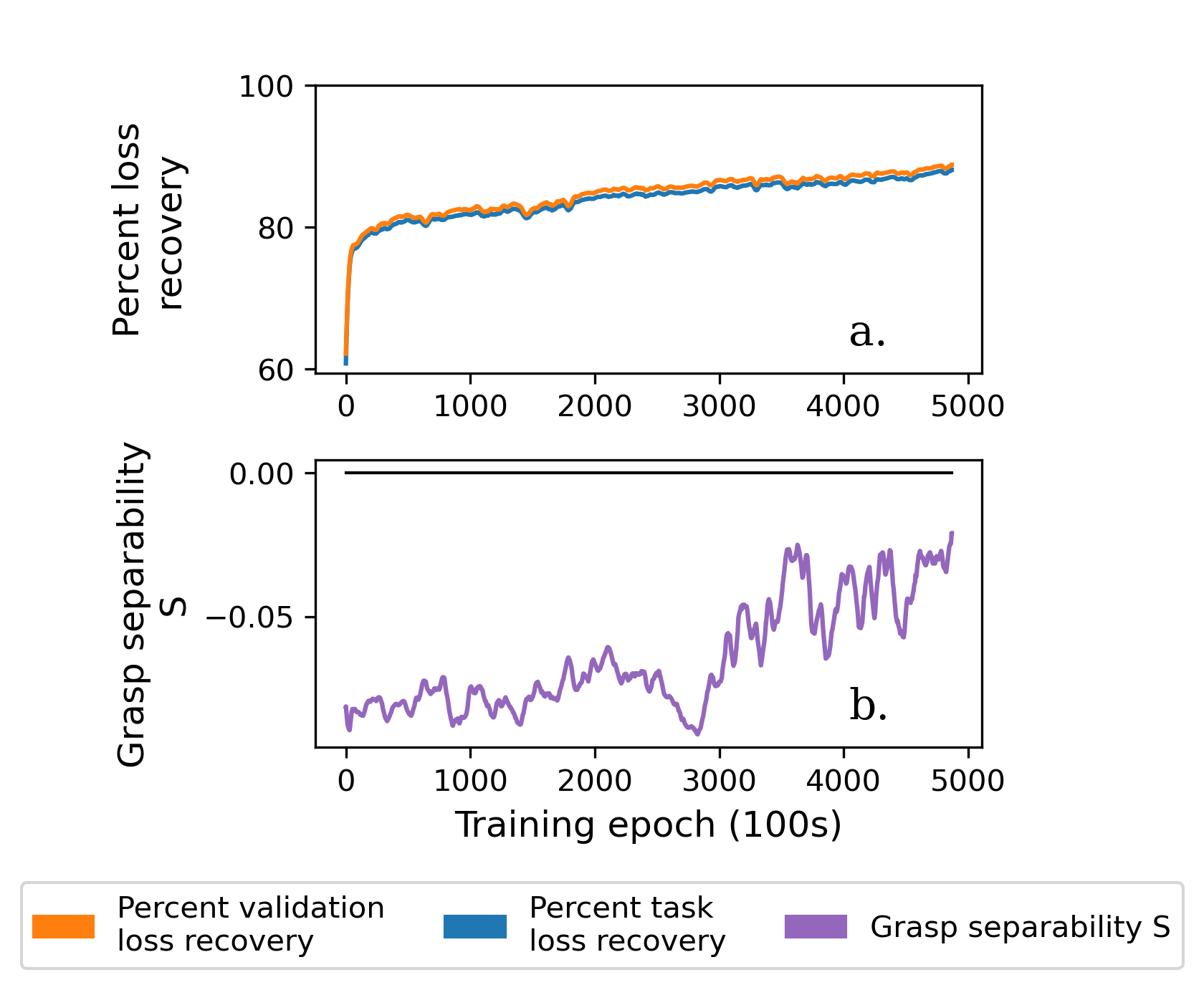}
\caption{Co-Processor Performance Results: Connection Lesion with Sensor Drift. a. 
         Percent loss recovery for the task and validation datasets. 0\% represents task loss equal to the loss for
         the lesioned cortical model, prior to any recovery. 100\% represents
         loss equal to that for the healthy cortical model. 
         b. Grasp separability metric (S) which reflects the co-processor's ability
         to tailor grasps to specific objects. A negative value indicates the grasps are less
         differentiated between object shapes. Due to sensor drift, the co-processor
         is much slower to differentiate objects and the trend exhibits 
         higher noise compared to other experiments.}
\label{fig:results_drift}
\end{figure}

As seen in Fig.~\ref{fig:results_drift}, the co-processor was able to quickly
recover the reach-to-grasp part of the task, and then gradually learned to
condition the grasp on the object information.\footnote{Note that we relaxed our 250k epoch
         stopping criterion for these results because sensor drift causes a longer
         convergence time.} Because the recording model is non-stationary
in this experiment, the object class separation exhibits far greater epoch-to-epoch
variability in the later epochs. The co-processor presumably becomes more
reliant on upstream visual information over time, as it learns to leverage this information 
in order to differentiate object shapes.

As expected, training efficiency was decreased relative to the experiment without sensor
drift: with the sensor drift, 282K training epochs were needed to reach $85\%$ recovery
(Fig.~\ref{fig:results_drift}a) while without sensor drift, only 93K were required
(Fig.~\ref{fig:results_con}a).

\section{Discussion and Future Work}
\label{sec:discussion}

We present here a first-of-its-kind demonstration of a novel design for neural co-processors,
showing in simulation crucial training and design methodologies that lay the groundwork for
co-processors to be demonstrated \textit{in vivo} in the future. Our results show that a deep
learning-based co-processor network (CPN) can learn neural stimulation policies that improve
performance of an external task after simulated lesions in different parts of a cortical model.
The training of the CPN relies on the use of a second neural network, called an emulator network (EN), 
which learns to approximate the function mapping stimulation and neural activity to task performance.
Our experiments revealed the effects of different parameter choices and training paradigms for CPNs and
ENs, providing new insights for future {\em in vivo} studies of neural co-processors.

Our co-processor design adapted well to a variety of simulated lesion types, reducing task loss
$75-90\%$ across our various experiments. The co-processor successfully adapted to
the long-running dynamics of the simulated cortical model, as well as the long-range effects of
stimulation. In some experiments, we required the co-processor to additionally adapt to non-stationarity in
the neural circuit, which was actively changing at the same time the co-processor was learning. In one
experiment, it also adapted to a simulated brain which had already undergone some amount of natural recovery
after a lesion. In this case, the co-processor successfully identified the information upstream from the 
lesion which was necessary to stimulate the motor cortex (M1) module downstream. In the final experiment,
the co-processor also successfully adapted to sensor drift simulated by a non-stationary recording model.

An important question that needs to be addressed is what can and cannot be inferred from simulation studies
such as ours. Today, predicting neural responses to stimulation over long time scales and complex stimulation
patterns remains a difficult problem, implying targeted control of neural activity is also difficult. Our
simulation results  do not constitute evidence that our method will, for example, immediately translate to
restoration of fine-grained control for grasping in a stroke patient. One clear challenge is the sheer
dimensionality of the problem. There exists a mismatch between the dimensionality of the brain and current
sensor and stimulator technologies. We are also severely limited in the amount of training data which can
reasonably be collected to train a closed-loop neural controller. We argue that while this fact clearly
creates a challenge for learning-based closed-loop stimulation, the insights we have gained through our
simulation studies are likely to be useful for {\em in vivo} testing of neural co-processors. 

As mentioned in the Background section~\ref{sec:background}, it is important to test 
the feasibility of applying a complex AI method such as a neural co-processor prior to testing it
\textit{in vivo}. Our simulations allowed us to design and test training methods around which
future \textit{in vivo} experiments can be developed. Additionally, iterating over different design
choices in simulation minimizes the use of animal experimentation, supporting a commonly-accepted
maxim regarding animal welfare in experimentation \cite{nrc.care}.

Our simulation studies required the co-processor to contend with issues of long-running dynamics and
non-stationarity, which are likely to be key issues facing real-world {\em in vivo} deployments. We designed
our training algorithm to contend with dimensionality by training the CPN with an EN that learns the effects
of stimulation. The training algorithm adapts to non-stationarity by regularly updating the EN, allowing it
to adapt to changes in the brain, in addition to ``following'' the CPN through stimulation parameter space.
In the next section, we briefly explore the possibility of further addressing the dimensionality problem by
using neural co-processors for optimizing low-dimensional neural correlates of task performance. 

\subsection{Co-processors for optimizing low-dimensional neural correlates of task performance}
Though our simulation involves a neural co-processor directly optimizing stimulation for
external task performance, the co-processor concept also generalizes to optimization for
\textit{neural correlates} (NCs) of task performance. We present this concept here to illustrate
the generality of our approach, and to provide a possible path towards tackling the dimensionality issue.
NCs are defined over measured brain activity and can be correlated with the desired patient outcome measures
such as improved dexterity. If such a correlate additionally possesses a causal relationship to the patient
outcome, i.e., if driving the NC through stimulation towards particular regimes also improves patient outcome,
then it may serve as a co-processor's optimization target. For example, Khanna et al. \cite{khanna.openloop}
suggest that ``changes in task-related ensemble firing are linked to improvements in dexterity.''
The authors then provide a measurement of ensemble co-firing which constitutes a candidate NC of dexterity
and which may potentially be driven through stimulation.

\begin{figure}[h]
	\centering
	\includegraphics[width=0.88\textwidth]{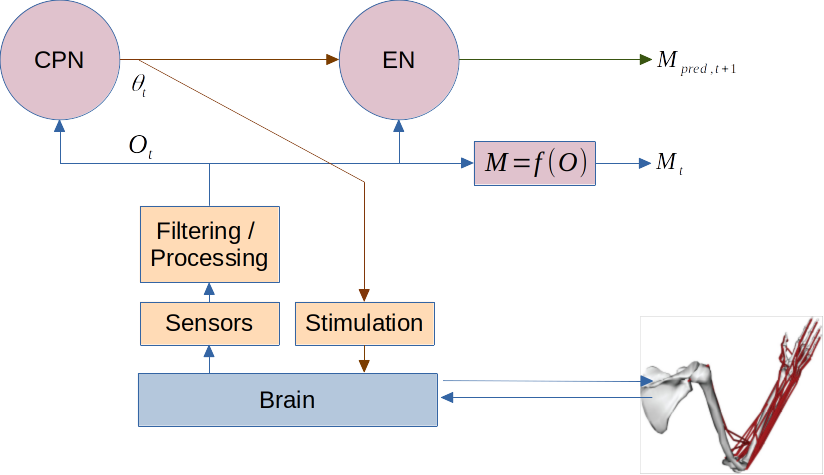}
	\caption{\textbf{A co-processor leveraging a neural correlate of dexterity rather
    than external (e.g., muscle) measurements.} Here, a measure $M$, serving as a neural correlate of
    dexterity, is calculated based on observed brain activity $O$. The emulator network (EN) learns
    to predict the future values of $M$ as a function of $O$ and stimulation parameters
     $\theta$. The co-processor network (CPN) maps the observed brain activity $O$ onto stimulation
    parameters. The CPN is trained using the co-processor approach presented in this article with the
    objective of maximizing the dexterity measure $M$ rather than minimizing a task loss defined in terms
    of grasping motions or muscle velocities.}
	\label{fig:correlate}
\end{figure}

A co-processor optimizing for an NCoD may provide a plausible path towards \textit{in vivo}
application. Driving a wide variety of grasping behaviors by altering the firing rates of a small number of
individual neurons or multiunits may be infeasible in practice, and may not generalize well beyond
the set of training examples. On the other hand, optimizing a neural correlate may generalize across a wide
variety of behaviors. Additionally, such an optimization problem may be lower dimensional and require less
training data in practice. We illustrate a design for such a co-processor in Fig.~\ref{fig:correlate}.


\subsection{Future work}
Identification of neural correlates such as NCoDs for different behaviors remains an open problem whose
solution may enable efficient types of future co-processors. Similar to Khanna et al.\ \cite{khanna.openloop},
Heimbuch et al.\ recently presented work suggesting that there exist low-dimensional neural correlates of
dexterity which can be measured during and after stroke recovery \cite{heimbuch.sfn}. They further hypothesize
that neural stimulation could drive brain activity in a stroke patient towards maximization of those metrics,
and that doing so would cause an improvement in dexterity. A co-processor could allow us to test
these hypotheses by attempting to learn a CPN stimulation policy which optimizes the candidate
NCoDs. Future work will involve collaborations with experimental neuroscientists to identify candidate NCs of
desirable behavior, and to perform {\em in vivo} experiments involving co-processors that seek to optimize
these NCs.

Additional future work will explore reinforcement learning (RL) approaches to training co-processors: the
neural co-processor we explored in this article already includes a ``policy network'' (the CPN) and a
``world model'' (the EN). The proposed framework is therefore well-suited to model-based reinforcement
learning algorithms such as model-based Actor-Critic learning. We will explore optimizing reward functions
that involve not only symptom relief, but also minimizing energy use by attaching a cost to stimulation.

One additional challenge remaining with our current co-processor design is data efficiency.
On the whole, the co-processor quickly improved task performance, but required orders of
magnitude more training examples to achieve its highest levels of performance. Even
moderate amounts of recovery may be valuable to a user, but nevertheless we consider
data efficiency to remain a problem. Due to limits on patient fatigue, time, implant
battery life, and other concerns, it is not plausible to expect a learning algorithm
to have access to unlimited amounts of training data. As a result, it is necessary to
make efficient use of the data we can acquire. We believe there exist at least four
mitigation strategies for this:

\begin{itemize}
	\item Retraining an existing EN, rather than repeatedly creating a new one. As
	      mentioned above, the former approach did not yield good results, but it remains
	      to be seen if this is a fundamental problem with our training method or
	    a peculiarity of our simulation which we have yet to identify. This
	      remains a future area of inquiry.
	\item Making better use of what data we acquire. In this initial simulation, we do
	      not retain data beyond the present training epoch because non-stationarity requires us to
	      regularly discard data as it ages. We found that in the presence
	      of a rapidly learning CPN, and especially in our co-adaptive experiments,
	      data retention in fact caused training instability. However, in
	      practice, training epochs operate on the order of seconds, suggesting
	      that data could be retained and reused for multiple epochs. The ``speed'' of 
	      non-stationarity compared to the co-processor's learning is an
	      area worthy of future investigation.
	\item Matching the dimensionality of the stimulation parameters and the optimization
        target to the amount of available data. A stimulation paradigm with
	      a small number of controllable parameters that nevertheless allows
	      improvements in task performance will likely require less training data. Likewise,
        optimizing low-dimensional targets (e.g., NCoDs) may require less training
        data.  
    \item Sharing data across subjects and transfer learning. Recent work has shown that
        it is possible to train a deep neural network using data from multiple subjects and
        transfer that knowledge to decode neural data from new subjects \cite{peterson.transfer}.
        A similar technique could potentially be used to train ENs, reducing the amount of
        data needed from a new subject and improving data efficiency.
\end{itemize}

\section{Conclusion}
Using a simulated model of cortical circuits in the primate brain that are involved in grasping
behaviors, we demonstrated a training method based on deep learning for a closed-loop neural
stimulator called a ``neural co-processor'' \cite{rao.coproc,rao.braincoproc}. We created a
learning paradigm which allowed the co-processor to adapt to the distributed neural activity
in the simulated brain, as well as to the non-stationarity of the neural activity and the 
sensors, key properties which are likely necessary for future \textit{in vivo} applications.
We showed that the co-processor could be trained to restore grasping function after we applied
varying types of simulated lesions. Specifically, a neural co-processor can be trained using
deep learning through backpropagation with the help of two networks: a co-processor
network (CPN) that learns to generate stimulation patterns to optimize task performance, and an
emulator network (EN) that learns to predict the effects of stimulation. Though the results we presented 
involved optimization of an external task, we believe that neural co-processors will successfully
generalize to the optimization of neural correlates of health, successfully driving neural activity
towards target regimes which are known to correlate with positive clinical outcomes. Given the generality
of the framework, we expect neural co-processors to be applicable to a wide range of clinical applications
that require adaptive closed-loop neural stimulation for treating sensorimotor disorders and
neuropsychiatric conditions. 

\section*{Code and data availability}
Code including analysis code used to
generate figures can be found at
\url{https://github.com/mmattb/coproc-poc}.
Data is available upon reasonable request to the authors.

\section*{Conflict of interest}
The authors of this work are not aware of any conflicts of interest
related to it.

\section*{Acknowledgements}
This work was supported by a Weill Neurohub Investigator grant, a CJ and Elizabeth Hwang
endowed professorship (RPNR), National Science Foundation (NSF) grant no.\ EEC-1028725 and NSF EFRI
grant no.\ 2223495. The authors would like to thank Karunesh Ganguly (UC San Francisco), Priya Khanna
(UC Berkeley), Anca Dragan (UC Berkeley), Justin Ong, Ian Heimbuch, and Luciano De La Iglesia for
discussions and insights related to this work.

\section*{References}
\bibliographystyle{iopart-num}
\bibliography{refs}

\newpage

\section{Supplementary Materials}
\subsection{Details of the simulated task}
\label{sup:michaelstask}

The reach-to-grasp task on which the mRNN cortical model was trained involves multiple phases, which can be identified by the
mRNN's inputs. Initially, the NHP subject sat in a dark room, with only a red cue visible. After a short period,
an object was presented to the subject visually, after which the overhead light was turned off and once again
only the red cue was visible. After another brief period, the red cue was turned off, which cued the subject
to perform a grasp and hold of the object (still in the dark). This task is presented in more detail in the
work of Sussillo \cite{sussillo.mrnn} and Michaels et al. \cite{michaels.mrnn}.

The inputs to the mRNN network encode the task phase information. It consists of a twenty dimensional vector
of visual features, which are presented only to the input (AIP) end of the network, and a 1D hold signal,
which is presented to all neurons across the network. The authors of Michaels et al. \cite{michaels.mrnn}
derived the hold signal from the grasp data, and it represents a point in time shortly before grasping began.
As explained above, the visual features are drawn from VGGNet \cite{simonyan.vgg}, based on rendered images.
The images consist of some combination of the red cue light and the object image. Note that the visual input
vector and hold signal are not observed directly by the co-processor, but rather indirectly via brain
recordings.

\begin{figure}[h!]
\centering
\includegraphics[scale=0.7]{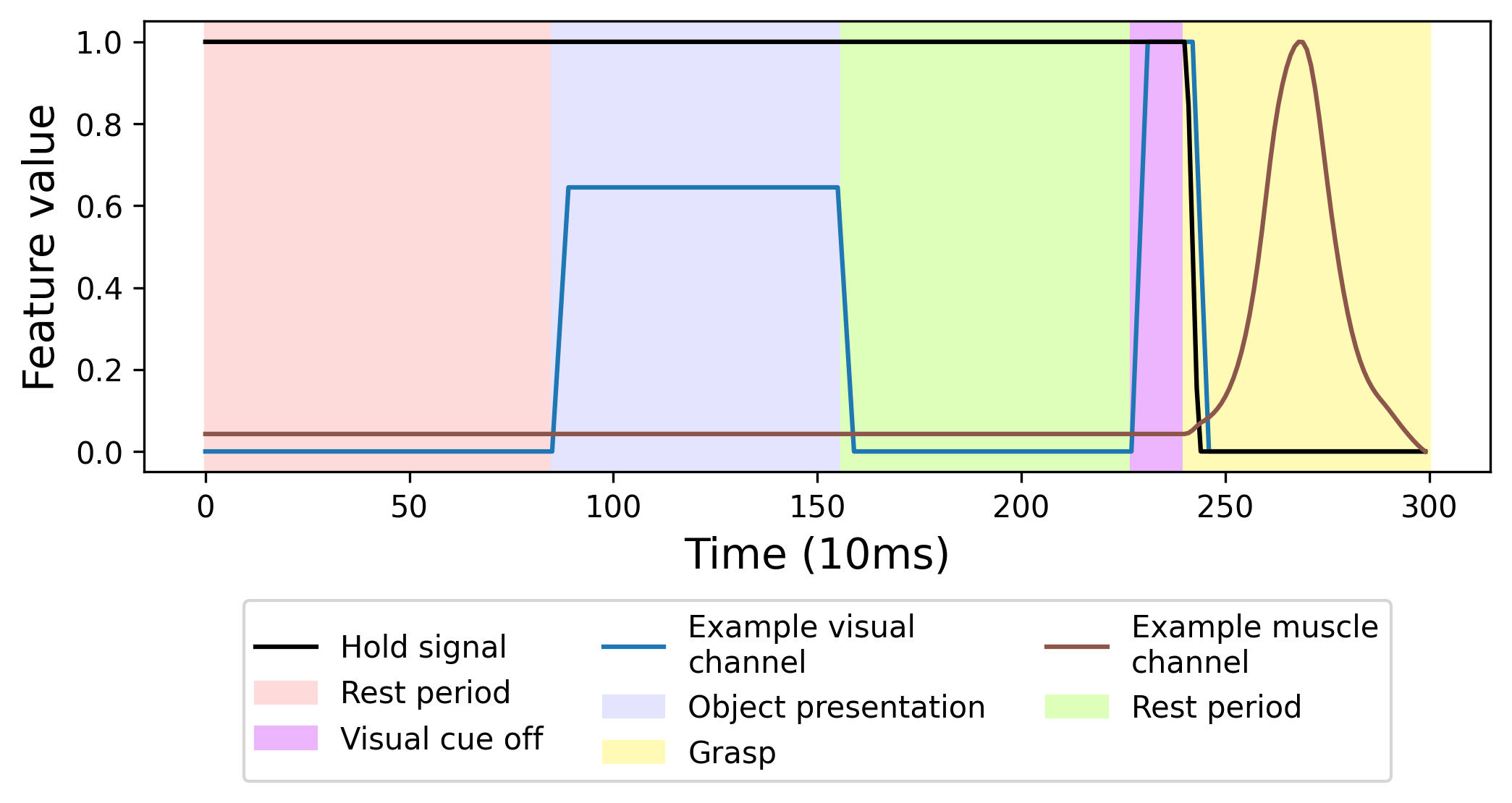}
\caption{Timeline of a Single Trial}
\label{fig:trial_timeline}
\end{figure}

The phases of the task are depicted in Table \ref{tab:phases}, and the timeline of a single trial is
depicted in Fig. \ref{fig:trial_timeline}.

\begin{table}[h!]
\begin{tabular}{|l|l|l|l|}
\hline
\textbf{Description}                                                                                                               & \textbf{Cue} & \textbf{\begin{tabular}[c]{@{}l@{}}Object\\ image\end{tabular}} & \textbf{\begin{tabular}[c]{@{}l@{}}Hold\\ signal\end{tabular}} \\ \hline
A rest period, during which a red cue light is presented                                                                           & On           & Off                                                             & On                                                             \\ \hline
\begin{tabular}[c]{@{}l@{}}The object presentation period, during which the\\ image of the object and cue are visible\end{tabular} & On           & On                                                              & On                                                             \\ \hline
A rest period                                                                                                                      & On           & Off                                                             & On                                                             \\ \hline
Cue is turned off                                                                                                                  & Off          & Off                                                             & On                                                             \\ \hline
Subject performs grasp                                                                                                             & On           & Off                                                             & Off                                                            \\ \hline
\end{tabular}
\caption{\label{tab:phases} \textbf{Phases of the reach to grasp task.} During each phase, a visual stimulus as well
         as a binary hold signal are presented. The visual stimulus encodes an image, consisting of a red cue light
         and/or an image of an object.}
\end{table}

\subsection{Training the cortical model for grasping}
\label{sup:michaelstraining}

For our simulation, we leverage a pre-trained mRNN cortical model, provided by the
lead author of Michaels et al. \cite{michaels.mrnn} and used with their permission.
The training method for that model is based on prior work by the same authors, and
others \cite{sussillo.mrnn}.

Training involved the use of Hessian-free optimization \cite{martens.hessianfree},
rather than the more common first-order stochastic gradient descent. Additionally, an L2 firing
rate regularization, and an L2 input and output model weight regularization were used.
These regularizations are believed to result in more biologically plausible RNNs \cite{kao.sim}.

The authors varied the model's nonlinearities, regularization weights, and
inter-module sparsities. Among those, we chose one model arbitrarily for our study. Specifically,
the model we chose was structured, and trained using:

\begin{itemize}
    \item The rectified hyperbolic tangent nonlinearity
    \item $1\mathrm{e}{-1}$ Inter-module sparsity
    \item $1\mathrm{e}{-3}$ L2 firing rate regularization
    \item $1\mathrm{e}{-5}$ L2 weight regularization
\end{itemize}

\subsection{EN/CPN Interleaving}
\label{sup:encpninter}
We train a CPN until an EN is no longer useful. At that point we train a new EN. We deem an EN
as no longer useful if one of the following predicates becomes true:

\begin{itemize}
	\item The EN prediction loss is greater than $\min(6\mathrm{e}{-4}, L/10)$, where $L$ is the most recent
	      task loss. EN prediction loss is an MSE loss between the EN's prediction of muscle
	      velocities, and the actual muscle velocities output by the grasping model network.
	\item Task loss increased across 15 of the previous 30 training epochs.
	\item 100 CPN training epochs have elapsed.
\end{itemize}

Likewise, the EN training period ends when its prediction loss $L$ on the validation data set drops
below a threshold of $\max(3\mathrm{e}{-4}, L/50)$.

\subsection{Sensor drift}
\label{sup:drift}
To simulate sensor drift, we add a vector to a bias term introduced to the recording function.
We draw the elements of the vector from a zero-mean Gaussian distribution, with variance based
on the mean value of the recording function from prior connection-lesion experiments. That
allows us to put it into a reasonable range, where it is effective but not extreme. We attempted
the experiment with several values of the variance, and found the results to be principally the same:
the co-processor eventually learned, but at a rate slower than otherwise. We present results for
a variance of $1.5\mathrm{e}{-3}$, which is $2\%$ of the mean recording value.

\subsection{Passthrough reference model}
\label{sup:passthrough}

To understand the effect of the recording and stimulation functions on the co-processor's performance, 
we repeated our experiment with the F5-M1 connection lesion and coadaptation using passthrough recording
and stimulation functions. ``Passthrough'' here means that the recording and stimulation functions have
dimensionality equal to the number of neurons we are recording from or stimulating, respectively, and that
there is no temporal or spatial smoothing used. That allows the co-processor to directly observe the hidden
state of the AIP and F5 neurons in the simulated grasping circuit, and to directly influence the hidden state
of the neurons in M1 it stimulates.

For this passthrough version of the experiment, we additionally increased the number of artifical neurons
making up the CPN and EN, to account for the additional dimensionality of the inputs and outputs. Specifically,
we increased the CPN from 61 to 200 neurons, and the EN from 87 to 351. We measured whether the increase in
neurons alone caused the different results we see here, and found it did not; we omit those results for
brevity.

\begin{figure}[h!]
\centering
\includegraphics[scale=0.7]{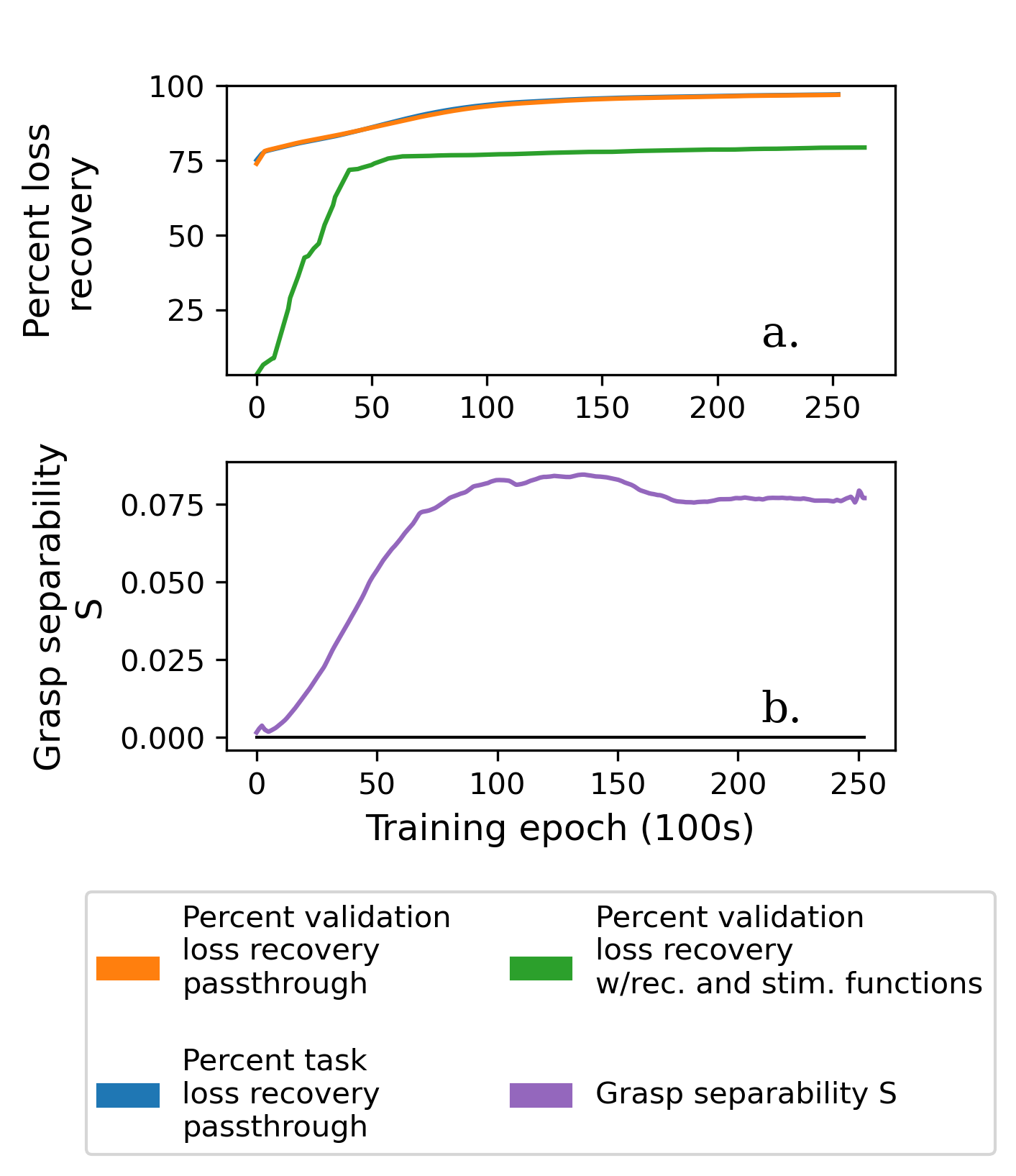}
\caption{\textbf{Training Results: passthrough recording and stimulation functions.} The significantly faster
         training and higher recovery suggest the recording and stimulation functions make the task
         more difficult for the co-processor.}
\label{fig:results_passthrough}
\end{figure}

In Fig. \ref{fig:results_passthrough} we see that the co-processor performed significantly better with
the passthrough functions. It learned faster, and to a higher recovery level: $97\%$ compared with $90\%$.
Note: we truncate the results to the point we stopped the passthrough experiment. The reader can find the
final results in Supplementary Materials \ref{sup:results}.

\subsection{Effect of observability on the recording function and training results}
\label{sup:observability}

\begin{figure}[h!]
\centering
\includegraphics[scale=0.7]{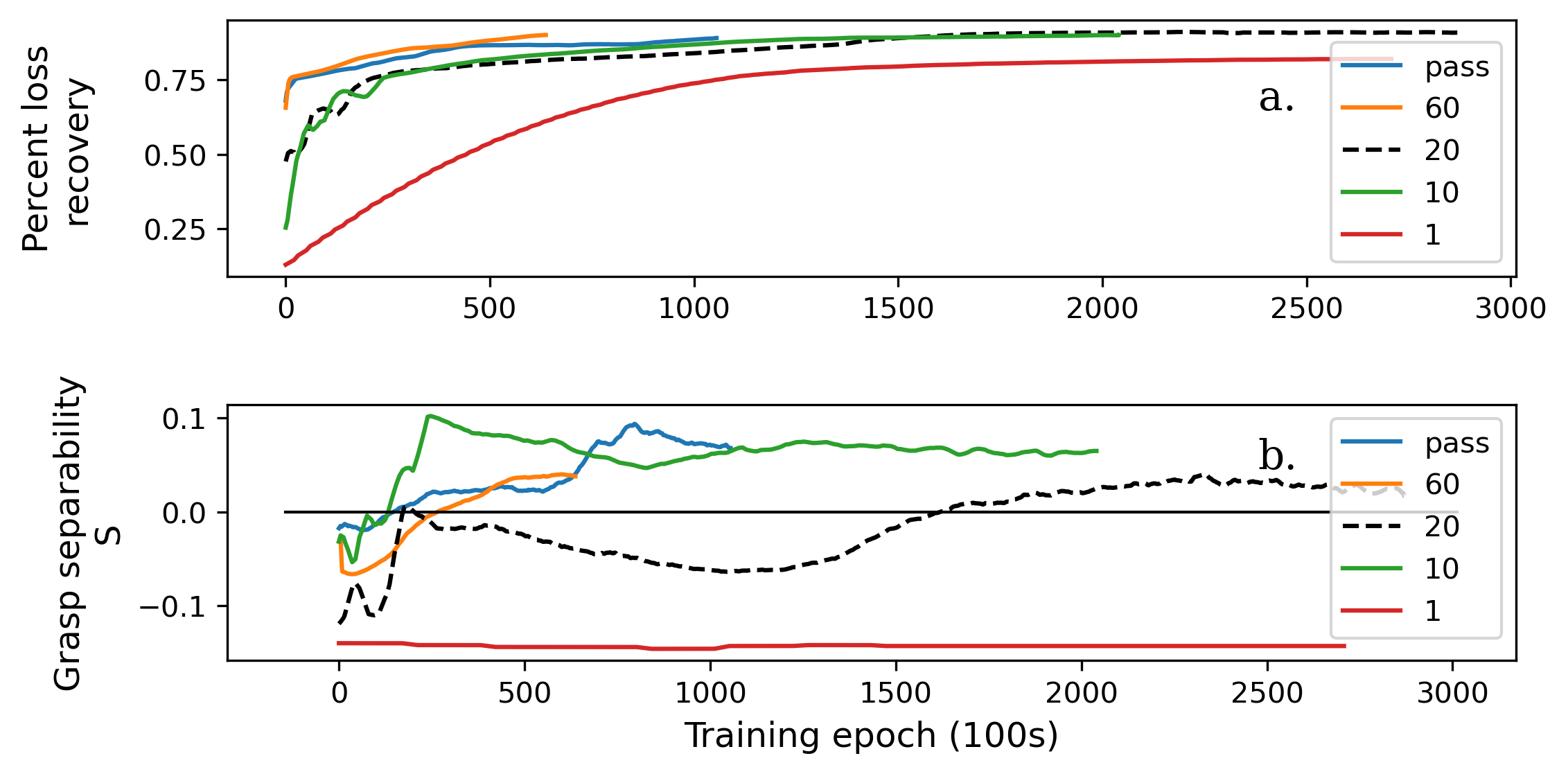}
\caption{\textbf{Training Results: varying observability of the brain.}
         (a) Increasing observability of the brain leads to faster learning. Here
         we show results with a single simulated recording electrode in each module
         ``1'', compared with ``20'' electrodes for the results reported in the main
         portion of this paper, and compared with other electrode counts. ``pass''
         refers to a passthrough recording function where every simulated neuron was
         directly measured. Note that learning is generally the same as observability
         increases, though perhaps somewhat faster. We did not hypothesis test that effect.
         In all cases except ``1'', learning reached a $90\%$ recovery threshold where
         we stop training. (b) As in our connection-based lesion experiment above
         (\ref{sec:results_con}), class separation initially varied, then trended towards the
         separation exhibited by the healthy mRNN network. However, in
         the single electrode experiment ``1'', class separation steadily decreased, suggesting
         that the information necessary to perform class separation cannot be determined from
         the information available to the single electrode.}
\label{fig:results_observability}
\end{figure}

To study the effect of brain observability on the co-processor's training behavior
we repeated the co-adaptive connection lesion experiment (\ref{sec:results_con})
with varying recording functions. Here we varied the number of simulated electrodes
in each mRNN module, and additionally performed an experiment with a passthrough recording
function, as described in the prior section. Specifically, our original experiment used
$20$ electrodes per module. Our additional experiments include $1$, $10$, and $60$ electrodes
per module, and the passthrough recording function. This allows us to explore any
gradient in the co-processor training results with respect to observability of the 
brain. Additionally, we reduced the variance of the Gaussian governing
the sensor model, such that an experiment with e.g. $1$ electrode per
module measures effectively only 1-2 neurons per module. We held the number of
neurons making up the CPN and EN fixed across these variations to ensure we
are examining the effects of the recording function alone. We continued to stop
learning at the point of $90\%$ recovery or a slow rate of recovery. In Fig.
\ref{fig:results_observability} we present the results.

We did not experiment with the placement of the electrodes, i.e. varying their
location among the neurons. It wasn't necessary to do so for the purpose of this
experiment. However, it's conceivable that results could improve considerably if we
happened to observe a small handful of key neurons.

Overall, learning is effectively the same regardless of the recording 
function, up to the point that only a small number of neurons can be 
observed. In all cases except the case of a single electrode, the co-processor
learned to improve task performance to the $90\%$ recovery point. Some variation
in learning efficiency occurred, and perhaps notably the experiments with $1$ and
$10$ electrodes observed the lowest two rates of recovery. The passthrough 
function and $60$ electrode model observed the fastest recovery rates.
Without additional computational power it remains difficult to hypothesis
test the effect of electrode count against learning efficiency, but the
results suggest that small variations in electrode count away from the
count of $20$ used in our main experiments would not cause drastically
different results.

Class separation behavior remained largely the same across these variations.
The one exception is the experiment involving only $1$ electrode per module.
In that case, the co-processor learned to treat the object classes
similarly. That suggests that the classes could not be differentiated by
the small amount of information attainable from a single electrode. In this
case, the co-processor reduced towards a simpler closed-loop stimulator
which modulates its behavior entirely on the user's volition to initiate
movement, due to the hold signal being observable from every neuron.

\subsection{Stopping criteria}
\label{sup:stopping_criteria}

Both stopping criteria affected the results presented above. We cannot economically perform
statistical analysis to show e.g. the percentage of runs of each experiment type which
are stopped by one criteria versus the other. However, in Table \ref{tab:stopping_criteria}
we present the stopping criterion which caused the end of each experiment whose results
we presented herein.

We used an alternate stopping condition for the sensor drift experiment, since recovery improved
more slowly in that experiment due to the sensor drift. We allowed the experiment to run until
roughly 500k epochs - essentially double the 250k epoch criterion.

We also used an alternate stopping condition for the passthrough functions experiment. This
was due to the computational expense: running the models at full resolution required far
greater run time. As a result, we stopped this experiment once $97\%$ recovery on the
validation data was achieved.

\begin{table}[h]
\centering
\begin{tabular}{|l|l|l|l|}
\hline
\multicolumn{1}{|c|}{\textbf{Experiment}} & \multicolumn{1}{c|}{\textbf{\begin{tabular}[c]{@{}c@{}}Loss\\ change \%\\ (default: 0.1\%)\end{tabular}}} & \multicolumn{1}{c|}{\textbf{\begin{tabular}[c]{@{}c@{}}Run length\\ (default: 250k\\ epochs)\end{tabular}}} & \multicolumn{1}{c|}{\textbf{Notes}}                                                                       \\ \hline
AIP No-coadapt                            & X                                                                                                      &                                                                                                                &                                                                                                           \\ \hline
AIP Coadapt                               &                                                                                                        & X                                                                                                              &                                                                                                           \\ \hline
M1 No-coadapt                             &                                                                                                        & X                                                                                                              & \begin{tabular}[c]{@{}l@{}}Pct change criteria relaxed to\\show longer-run S trend\end{tabular}            \\ \hline
M1 Coadapt                                &                                                                                                        & X                                                                                                              &                                                                                                           \\ \hline
Con No-coadapt                            & X                                                                                                      &                                                                                                                & Loss chg threshold: 0.05\%                                                                                \\ \hline
Con Coadapt                               &                                                                                                        & X                                                                                                              & Loss chg threshold: 0.05\%                                                                                \\ \hline
Recovery                                  &                                                                                                        & X                                                                                                              &                                                                                                           \\ \hline
Sensor drift                              & X                                                                                                      &                                                                                                                & \begin{tabular}[c]{@{}l@{}}Allowed to run to $\sim$500k epochs,\\due to slower convergence.\end{tabular} \\ \hline
Passthrough                               &                                                                                                        &                                                                                                                & \begin{tabular}[c]{@{}l@{}}Stopped at 97\% recovery,\\due to computational expense.\end{tabular}         \\ \hline
\end{tabular}
\caption{\label{tab:stopping_criteria} \textbf{Stopping criterion which bound each experiment.}
         ``Coadapt'' refers to co-adaptation. ``Con'' refers to F5-M1 connection lesions.
         ``Recovery'' refers to the experiment involving recovery prior to co-processor training.}
\end{table}

\subsection{Prediction Accuracy of Stimulation Model (EN)}
\label{sup:pred_losses}

\begin{figure}[h!]
\centering
\includegraphics[scale=0.8]{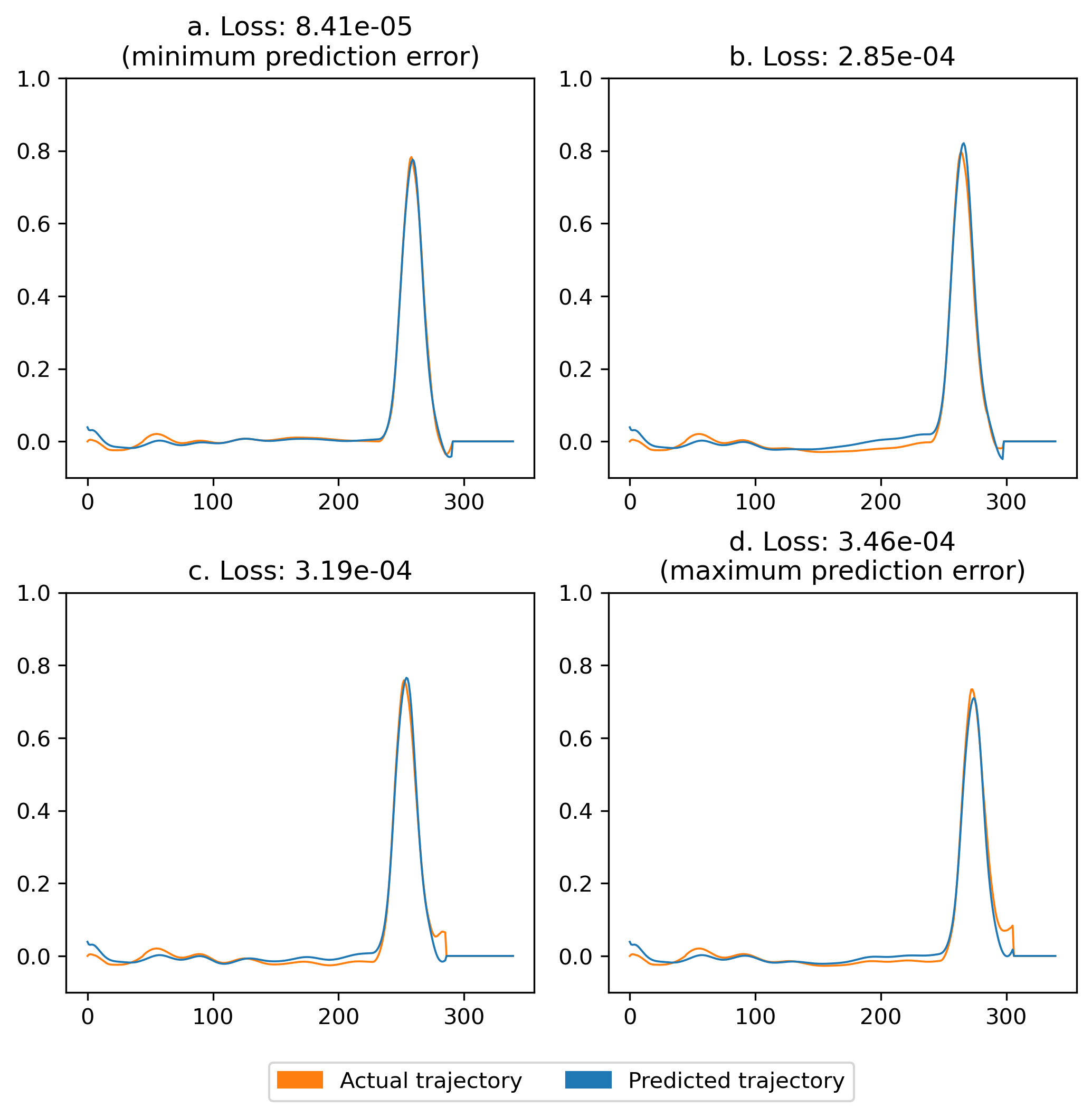}
\caption{\textbf{Prediction errors for four trials, on TMAJ muscle (retracts the shoulder).}
         The EN's predicted trajectory for the TMAJ muscle (blue), closely tracks the actual
         trajectory (orange). Sourced from the co-adaptive connection lesion experiment.
         (a) depicts the trial with minimum prediction MSE. (d) maximum prediction
         MSE. (b, c) two other arbitrary-chosen trials. Standard deviation across all trials:
         $1.91\mathrm{e}{-5}$.}
\label{fig:pred_loss_examples}
\end{figure}

As mentioned in Section~\ref{sec:en_training} above, we train our stimulation model, i.e. EN, to a high
level of predictive power before using it to train the CPN. We expire it and create a new one when training
performance begins to suffer, or prediction error significantly disimproves. To depict the EN's predictions
visually, Fig. \ref{fig:pred_loss_examples} shows example muscle velocities for a highly trained CPN and EN,
for one example output channel.

\begin{figure}
	\centering
	\begin{subfigure}[c]{0.48\textwidth}
		\centering
		\includegraphics[width=\textwidth]{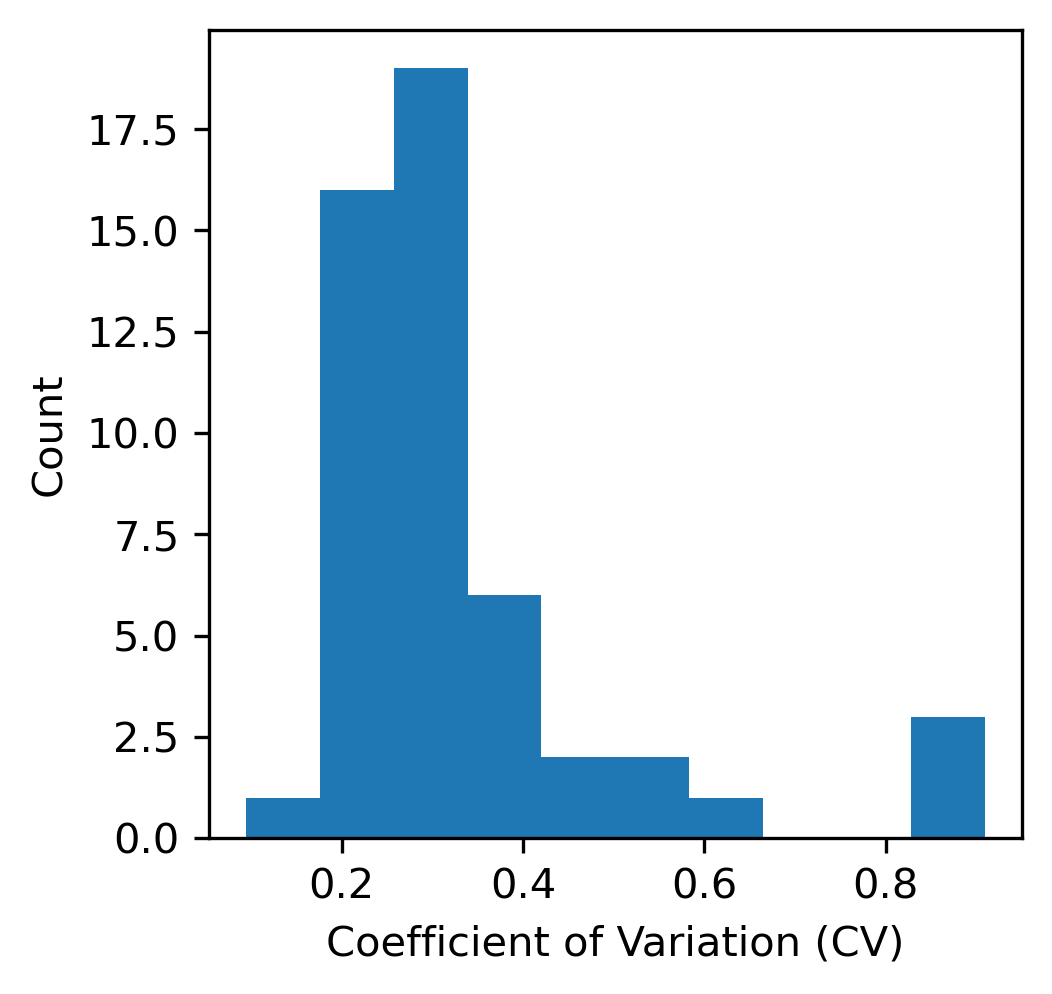}
		\caption{\textbf{Histogram of coefficients of variability (CVs) across prediction MSE
                 losses for each muscle channel.} Mean: $0.340$; i.e. per-channel loss
                 typically varies $\pm34.0\%$.}
	\end{subfigure}
	\hfill
	\begin{subfigure}[c]{0.48\textwidth}
		\centering
		\includegraphics[width=\textwidth]{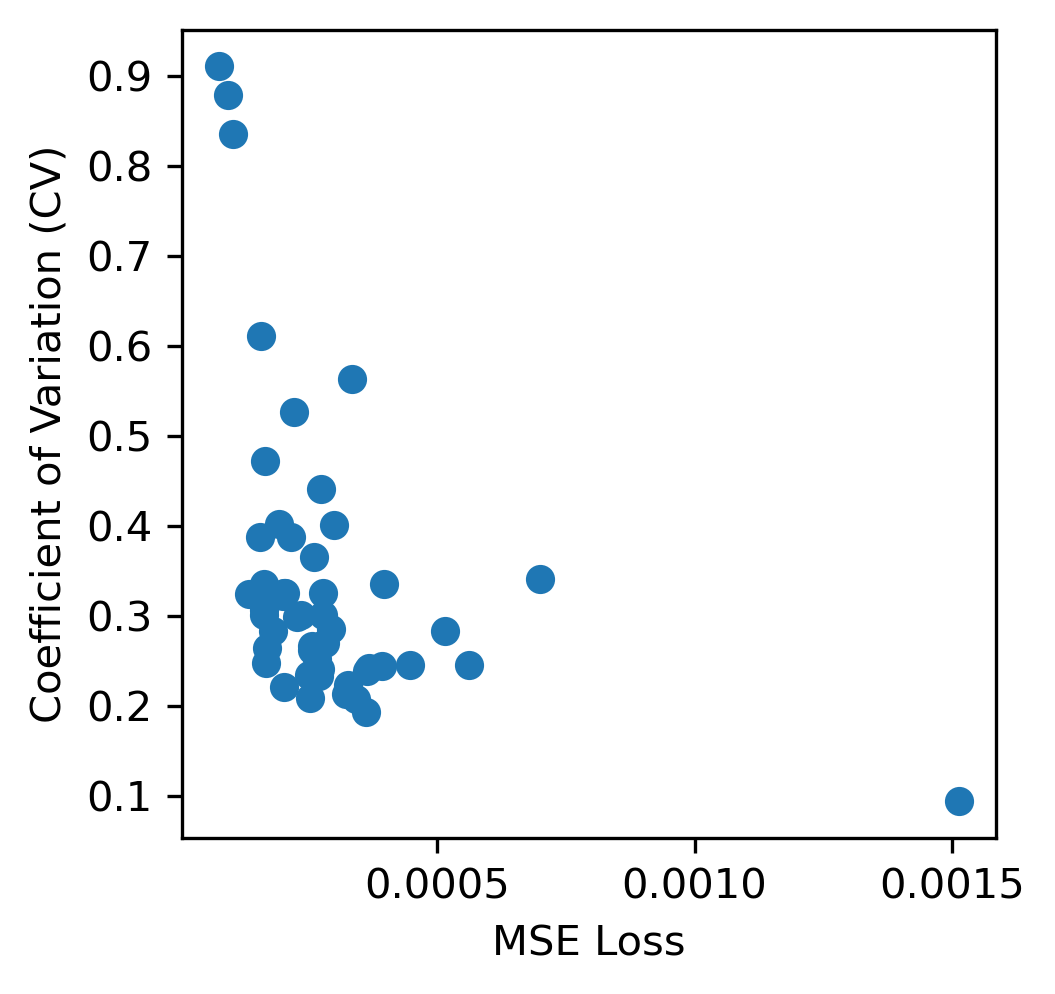}
		\caption{\textbf{Prediction MSE loss vs CV for each channel.} High CV extrema are from channels
                 with the lowest losses. i.e. the highest CV values in (a) are due to extreme
                 low losses, suggesting their CVs are driven by the low mean, rather than an
                 unusually high variability.}
	\end{subfigure}
	\hfill
\caption{\textbf{Prediction loss variability across muscle channels}}
\label{fig:pred_loss_cv}
\end{figure}

In Fig. \ref{fig:pred_loss_cv} we see the variability of prediction losses across muscle
channels. Prediction loss typically varies $\pm34.0\%$ for a given channel across all
trials. Variability is roughly consistent across channels, suggesting that the statistic
is not driven by extrema. However, there exist a handful of channels with high
coefficients of variation (CVs). Those appear to be the channels with the lowest overall
prediction losses, suggesting their CVs are driven by the low mean, rather than an unusually
high variability.

\subsection{Table of losses and recovery}
\label{sup:results}

\begin{table}[h]
\begin{tabular}{|l|l|l|l|l|l|}
\hline
\textbf{Experiment}  & \textbf{\begin{tabular}[c]{@{}l@{}}Lesioned\\ loss\end{tabular}} & \textbf{\begin{tabular}[c]{@{}l@{}}Min task\\ loss\end{tabular}} & \textbf{\begin{tabular}[c]{@{}l@{}}Min task val\\ loss\end{tabular}} & \textbf{\begin{tabular}[c]{@{}l@{}}Pct recov\end{tabular}} & \textbf{\begin{tabular}[c]{@{}l@{}}Pct recov\\ val\end{tabular}} \\ \hline
AIP No-coadapt & 0.004507                                                          & 0.003916                                                         & 0.003960                                                             & 15.01\%            & 13.89\%                                                          \\ \hline
AIP Coadapt    & 0.004507                                                          & 0.001575                                                         & 0.001502                                                             & 74.23\%            & 76.26\%                                                          \\ \hline
M1 No-coadapt  & 0.021136                                                          & 0.005765                                                         & 0.005604                                                             & 74.73\%            & 75.51\%                                                          \\ \hline
M1 Coadapt     & 0.021136                                                          & 0.004383                                                         & 0.004385                                                             & 81.39\%            & 81.41\%                                                          \\ \hline
Con No-coadapt & 0.020719                                                          & 0.002540                                                         & 0.002490                                                             & 90.21\%            & 90.45\%                                                          \\ \hline
Con Coadapt    & 0.020719                                                          & 0.002350                                                         & 0.002521                                                             & 91.15\%            & 90.30\%                                                          \\ \hline
Recovery*      & 0.003834                                                          & 0.003136                                                         & 0.003099                                                             & 18.20\%            & 19.17\%                                                          \\ \hline
Sensor drift   & 0.020719                                                          & 0.002966                                                        & 0.002825                                                              & 88.13\%            & 88.83\%                                                          \\ \hline
Passthrough    & 0.020719                                                          & 0.001095                                                        & 0.001132                                                              & 97.00\%            & 96.83\%                                                          \\ \hline
\end{tabular}
\caption{\label{tab:results}Losses and recovery. *Lesioned loss and percent recoveries based on the post-recovery values.
         ``Coadapt'' refers to co-adaptation. ``Con'' refers to F5-M1 connection lesions. ``Recovery'' refers to the
         experiment involving recovery prior to co-processor training. For further exploration of the results, see Results.}
\end{table}
\end{document}